\documentclass[pra,aps,amsmath,amssymb,amsfonts,twocolumn,nofootinbib,longbibliography,superscriptaddress]{revtex4-2}
\usepackage{amssymb}
\usepackage{bm,mathrsfs}
\usepackage{graphicx}
\usepackage{epsfig}
\usepackage{amsmath,bbm}
\usepackage{amsfonts,amssymb}
\usepackage{times}
\usepackage{verbatim}
\usepackage[sort&compress]{natbib}
\usepackage{amsmath}
\usepackage{bm}
\usepackage{float}
\usepackage{braket}
\usepackage{booktabs}
\usepackage{xcolor}

\allowdisplaybreaks[4]
\usepackage[colorlinks,breaklinks,linkcolor=blue,anchorcolor=blue,citecolor=blue,urlcolor=blue]{hyperref}
\begin{document}

\title{Accelerated Rydberg electromagnetically induced transparency quantum memory via shortcuts to adiabaticity}

\author{Y. Wei}
\affiliation{Center for Quantum Sciences and School of Physics, Northeast Normal University, Changchun 130024, China}

\author{Changcheng Li}
\affiliation{State Key Laboratory of Quantum Optics Technologies and Devices, Institute of Laser Spectroscopy, Shanxi University, Taiyuan 030006, China}

\author{Y. M. Liu}
\affiliation{Center for Quantum Sciences and School of Physics, Northeast Normal University, Changchun 130024, China}

\author{Yuechun Jiao}
\affiliation{State Key Laboratory of Quantum Optics Technologies and Devices, Institute of Laser Spectroscopy, Shanxi University, Taiyuan 030006, China}

\author{Weibin Li}
\affiliation{School of Physics and Astronomy, and Centre for the Mathematics and Theoretical Physics of Quantum Non-equilibrium Systems, The University of Nottingham, Nottingham NG7 2RD, United Kingdom}

\author{X. Q. Shao}
\email{Contact author: xqshao@nenu.edu.cn}
\affiliation{Center for Quantum Sciences and School of Physics, Northeast Normal University, Changchun 130024, China}
\affiliation{Institute of Quantum Science and Technology, Yanbian University, Yanji 133002, China}

\begin{abstract}
Electromagnetically induced transparency (EIT) enables coherent light-matter storage, forming the basis of photonic quantum memories that are essential for scalable quantum networks and distributed quantum computing.  However, accelerating the storage process violates the adiabatic condition, resulting in the excitation of the lossy intermediate state and a reduction in writing efficiency. We propose and numerically investigate a high-speed, high-fidelity quantum storage scheme by incorporating a shortcut-to-adiabaticity (STA) technique based on counter-diabatic (CD) driving. By introducing a precisely engineered auxiliary field into a conventional EIT system, our protocol significantly shortens the writing time beyond the conventional adiabatic limit while effectively suppressing the transient population of the lossy intermediate state. Furthermore, our scheme demonstrates strong flexibility in pulse design, remaining effective across different temporal profiles of both the control and signal fields. It also exhibits robustness against imperfections in the CD drive. Even with imperfect single-photon writing and non-ideal Rydberg blockade, the scheme retains clear advantages, maintaining high storage performance and overcoming the intrinsic speed-fidelity trade-off of traditional EIT protocols. These features pave the way for fast and robust quantum devices suitable for high-throughput quantum repeaters and advanced quantum information processing.
\end{abstract}

\maketitle

\section{Introduction}\label{sec1}
As a cornerstone for building large-scale quantum networks and distributed quantum computing, quantum memory serves as an essential component of quantum information technology~\cite{Lvovsky2009,Hedges2010,Heshami2016,li2016quantum,Distante2017,PhysRevLett.131.240801,Wei2024,PhysRevX.14.021018,kbwj-md9n,Teller2025}. Among the various physical implementations, atomic-ensemble-based schemes have attracted considerable attention owing to their excellent coherence properties and scalability. A key mechanism underlying these schemes is EIT~\cite{Lukin2001Controlling,PhysRevA.65.022314,PhysRevLett.102.170502,PhysRevLett.107.213601,PhysRevA.89.033839,Lei2022}, which enables the opening of a narrow transparency window in an otherwise opaque atomic medium through the introduction of a strong coupling field. Within this transparency window, photon pulses carrying quantum information can be coherently mapped onto the collective atomic excitation in the form of dark-state polaritons (DSP), thereby realizing high-fidelity optical storage~\cite{PhysRevLett.98.243602,Guo2019,PhysRevA.104.033714,PhysRevResearch.4.023024,PhysRevA.108.032618}. Such EIT-based quantum memories provide a coherent interface between flying photonic qubits and stationary atomic excitations, and play a central role in a variety of quantum information tasks, including single-photon generation~\cite{Zhou:12,PhysRevLett.110.103001,PhysRevLett.112.073901,Wang2019,PhysRevA.101.013421,PhysRevResearch.3.033287,Fan:23,PhysRevLett.131.133001}, quantum communication and networking~\cite{Busche2017,Thompson2017}, entanglement distribution~\cite{PhysRevLett.121.123603,Shi_2022,PhysRevLett.128.060502,PhysRevLett.133.233003}, and photonic quantum logic operations~\cite{Saffman_2016,PhysRevA.98.052324,Cotrufo:19,shi2022high,PhysRevA.105.042430,PhysRevA.111.022420}.

To enhance optical storage performance and realize strong photon–photon nonlinearities, the combination of EIT with Rydberg atoms has become a powerful platform in quantum optics and quantum information science~\cite{PhysRevA.92.053846,PhysRevX.5.031015,Weber2015,Letscher_2017,PhysRevResearch.2.043339,PhysRevX.12.021034,yang2022sequential,Kumlin_2023,10.1063/5.0211071,PhysRevLett.133.173604,cmpn-jfqr}. Owing to their large electric dipole moments, Rydberg atoms exhibit strong dipole–dipole or van der Waals interactions, leading to the Rydberg blockade effect~\cite{RevModPhys.82.2313,PhysRevLett.133.213601}. Under blockade conditions, a small atomic ensemble can absorb at most one photon and collectively form a Rydberg superatom, thereby inhibiting further photon absorption~\cite{peyronel2012quantum,PhysRevLett.125.073602,Yang:22,Srakaeew2023,7zjs-73qm}. This collective behavior markedly enhances the light–matter interaction, enabling Rydberg-EIT storage schemes to operate in the regime of single-photon nonlinear optics and high-efficiency quantum memories. In recent years, this mechanism has been experimentally exploited to demonstrate single-photon storage and nonclassical light manipulation, establishing Rydberg-EIT systems as a key resource for scalable quantum repeaters and quantum networks~\cite{doi:10.1126/sciadv.1600036,PhysRevA.97.043811,PhysRevLett.121.123605,PhysRevLett.123.140504,Ornelas-Huerta:20}.

Despite these advances, EIT-based optical storage relies on the adiabatic following of a dark state, a feature shared with related coherent control schemes such as stimulated Raman adiabatic passage (STIRAP)~\cite{RevModPhys.89.015006,RevModPhys.91.045001,PhysRevA.102.013706}. However, unlike STIRAP, which is primarily concerned with internal-state population transfer, EIT-based storage intrinsically involves the propagation and storage of optical fields in an extended medium. Furthermore, the adiabatic nature of EIT-based storage makes the process inherently slow, while also rendering it highly susceptible to decoherence and experimental noise. Although numerous counter-adiabatic schemes have been developed so far, most focus solely on the internal states of atoms, aiming to control only the internal degrees of freedom~\cite{Bason2012,PhysRevLett.110.240501,PhysRevA.90.022307,PhysRevA.91.032304}. In contrast, our goal is to assess the practical applicability of STA for realizing an atom–photon interface. To this end, in this work we employ STA, utilizing engineered CD driving to accelerate the writing of signal photons in Rydberg-superatom-based EIT storage~\cite{PhysRevLett.105.123003,PhysRevLett.111.100502,PhysRevA.89.033856,Du2016,PhysRevA.94.063411,Zhao2017Robust,PhysRevLett.126.023602,Wu2026}. Our results show that the CD field effectively suppresses the non-adiabatic excitation of the lossy intermediate state induced by rapid modulation, enabling high-fidelity transfer of photonic excitations into the collective Rydberg state on timescales far shorter than those required by conventional adiabatic protocols. By further analyzing the propagation dynamics of the signal field, we demonstrate that the proposed scheme significantly improves the writing efficiency compared to the case without CD driving. Moreover, the protocol exhibits strong robustness against variations in the input signal field and imperfections in the CD control. Even considering the potential imperfect conditions in the experiment, such as multiphoton inputs or imperfect Rydberg blockade, the scheme still maintains clear advantages in storage performance. Overall, our approach enables fast and efficient signal photon writing, providing a promising route toward high-speed and robust quantum memories based on Rydberg ensembles.

The remainder of this paper is structured as follows. In Sec.~\ref{sec2}, we present the theoretical model of the proposed scheme. In Sec.~\ref{sec3}, we discuss the results, including the dynamical evolution of the spin‑wave state, the propagation of the signal field, and the robustness of the scheme against experimental imperfections and control errors, as well as the influence of multiphoton inputs and imperfect Rydberg blockade. Finally, we summarize our findings in Sec.~\ref{sec4}.

\section{Model and methods}\label{sec2}

\subsection{The General N-Atom Hamiltonian}
\begin{figure}
	\centering
	\includegraphics[width=0.9\linewidth]{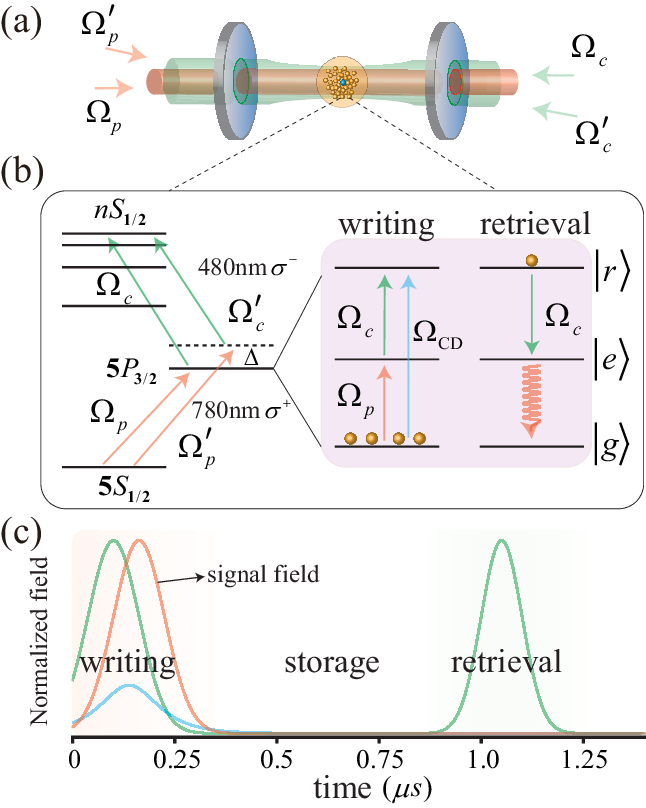}
	\caption{(a) An ensemble of cold $^{87}\text{Rb}$ atoms is confined in a dipole trap. (b) Relevant atomic level structure illustrating the EIT-based storage and retrieval processes, including the CD driving field. (c) Pulse sequences of the signal, control, and CD fields.}\label{fig1}
\end{figure}

We begin with the full Hamiltonian of a system consisting of $N$ three-level $^{87}\text{Rb}$ atoms, as schematically illustrated in Fig.~\hyperref[fig1]{1(a)}. The three relevant atomic levels are the ground state $\ket{g} \equiv |5S_{1/2},F=1,m_F=0\rangle$, the intermediate excited state $\ket{e} \equiv |5P_{3/2},F=2,m_F=1\rangle$, and the Rydberg state $\ket{r} \equiv |nS_{1/2},m_j=1/2\rangle $, whose level structure and optical couplings are shown in Fig.~\hyperref[fig1]{1(b)}. In the rotating frame, the total interaction Hamiltonian can be expressed as $ H = H_{\text{a}} + H_{\text{int}}$. The light-matter interaction is described by the Hamiltonian ${H}_{\text{a}}$ in the interaction picture ($\hbar=1$):
\begin{align}
H_{\text{a}} =  -\sum_{i=1}^{N} \Big[ g a \, e^{\mathbf{i} \mathbf{k}_p \cdot \mathbf{r}_i} \, {\sigma}_{eg}^{i} + \frac{\Omega_c(t)}{2} \, e^{\mathbf{i} \mathbf{k}_c \cdot \mathbf{r}_i} \, {\sigma}_{re}^{i}\Big] + \text{H.c.} 
\end{align}
Where ${\sigma}_{\mu \nu}^{i} = \ket{\mu}_i\bra{\nu}$ denotes the transition operator of the $i$th atom. The quantum probe field, described by the annihilation operator $a$, couples the $\ket{g} \leftrightarrow \ket{e}$ transition with a single-photon coupling strength $g = \wp_{ge} \sqrt{\omega_p / (2 \epsilon_0 V)}$, where $\wp_{ge}$ is the electric dipole matrix element, $\omega_p$ is the probe frequency, $\epsilon_0$ is the vacuum permittivity, and $V$ is the quantization volume. The classical control field drives the $\ket{e} \leftrightarrow \ket{r}$ transition with a time-dependent Rabi frequency $\Omega_c(t)$.

The strong, long-range van der Waals interactions between Rydberg atoms are described by:
\begin{align}
H_{\text{int}} = \sum_{i<j} U_{rr} {\sigma}_{rr}^{i} {\sigma}_{rr}^{j},
\end{align}
where $U_{rr} = {C_6}/{|\mathbf{r}_i - \mathbf{r}_j|^6}$, with $\mathbf{r}_i$ and $\mathbf{r}_j$ denoting the positions of the $i$th and $j$th atoms, respectively, and $C_6$ is the van der Waals dispersion coefficient for the $nS$ Rydberg state. In the Rydberg blockade regime, the system dynamics are restricted to the collective excitation manifold. We therefore adopt a semiclassical approximation by replacing the quantum probe field operator with its expectation value. This leads to an effective probe Rabi frequency $\Omega_p(t)=2g\langle {a}(t)\rangle$, which is collectively enhanced by a factor of $\sqrt{N}$ in the superatom basis. To streamline the theoretical analysis and numerical calculations, both the probe field $\Omega_p(t)$ and the control field $\Omega_c(t)$ are modeled as Gaussian pulses:
\begin{equation}
    \Omega_n(t) = \Omega_n 
\exp\!\left[-\frac{(t - t_n)^2}{2\sigma_n^2}\right],
\quad (n \in \{p,c\}),
\end{equation}
where $\Omega_n$ denotes the peak Rabi frequency, $t_n$ is the center time of the pulse, and $\sigma_n$ determines the standard deviation of the Gaussian envelope. The control pulse $\Omega_c(t)$ is temporally displaced relative to the probe pulse $\Omega_p(t)$ to satisfy the timing requirements for the storage and retrieval protocols.

\subsection{Rydberg Blockade and Collective States}
The interaction term ${H}_{\text{int}}$ is the source of the Rydberg blockade. The $U_{rr}$ is extremely large for nearby atoms, causing a significant energy shift that prevents the excitation of more than one atom to the Rydberg state within a blockade volume. Although fundamentally quantum in nature, the probe field is treated as an extremely weak signal within the semi-classical framework. In the context of single-photon storage, this exceptionally weak field is conventionally regarded as a single excitation. Consequently, the system's dynamics are constrained to the single-excitation subspace. This allows us to describe the N-atom ensemble using a set of collective states. We define the following basis states for the superatom, which properly account for the phase matching with the driving laser fields:
\begin{align}
    \ket{G} &= \ket{g_1, g_2, \dots, g_N},\notag\\
    \ket{E} &= \frac{1}{\sqrt{N}} \sum_{i=1}^{N} e^{\mathbf{i}\mathbf{k}_p \cdot \mathbf{r}_i} \ket{g_1, \dots, e_i, \dots, g_N},\notag\\
     \ket{R} &= \frac{1}{\sqrt{N}} \sum_{i=1}^{N} e^{\mathbf{i}(\mathbf{k}_p + \mathbf{k}_c) \cdot \mathbf{r}_i} \ket{g_1, \dots, r_i, \dots, g_N}.
\end{align}   
The phase factors, determined by the wave vectors of the probe ($\mathbf{k}_p$) and coupling ($\mathbf{k}_c$) fields, ensure constructive interference and efficient coupling to the light fields. In this collective basis, the light-matter coupling strength is enhanced by a factor of $\sqrt{N}$. Consequently, the system dynamics are described in the basis of collective atomic states $\{ \ket{G}, \ket{E}, \ket{R} \}$. The effective interaction Hamiltonian in the rotating frame is given by:
\begin{align}
    {H}_{\text{eff}}(t) = -\frac{1}{2}
    \begin{pmatrix}
    0 & \sqrt{N}\Omega_p(t) & 0 \\
    \sqrt{N}\Omega_p^*(t) & 0 & \Omega_c(t) \\
    0 & \Omega_c^*(t) & 0
    \end{pmatrix}.
\end{align}
Then, we solve the eigenvalue $H_{\text{eff}}\ket{\Psi} = E\ket{\Psi}$. By diagonalizing the effective Hamiltonian, we obtain its instantaneous eigenvalues and eigenstates. One eigenvalue is identically zero ($E_0 = 0$), which corresponds to a unique superposition known as the dark state. The other two eigenvalues are non-zero, given by $E_{\pm} = \pm \frac{1}{2} \Omega_{\text{Rabi}}(t)$. These non-zero eigenvalues correspond to a pair of bright states. The instantaneous dark state and bright states of the coupled atom–photon system are given by
\begin{align}
    \ket{\Psi_0(t)} = & \frac{1}{\Omega_{\text{Rabi}}(t)} (\Omega_c(t)\ket{G}- \sqrt{N} \Omega_p(t)\ket{R}),\notag\\
    \ket{\Psi_+(t)} = & \frac{1}{\sqrt{2}\Omega_{\text{Rabi}}(t)} (\sqrt{N} \Omega_p(t) \ket{G}\notag \\
    &+ \Omega_{\text{Rabi}}(t)\ket{E} + \Omega_c(t)\ket{R} ),\notag\\
    \ket{\Psi_-(t)} = & \frac{1}{\sqrt{2}\Omega_{\text{Rabi}}(t)} (\sqrt{N} \Omega_p(t)\ket{G}\notag \\
    &- \Omega_{\text{Rabi}}(t)\ket{E} + \Omega_c(t)\ket{R}),    
\end{align}
where $\Omega_{\text{Rabi}}(t)=\sqrt{\Omega_c(t)^2+ N\Omega_p(t)^2}$. The dark state $\ket{\Psi_0(t)}$, which is decoupled from the excited state $\ket{E}$, can therefore be interpreted as a DSP, corresponding to a coherent superposition of a photonic excitation and a collective Rydberg spin wave. Coherent storage and retrieval of the signal field can be achieved by adiabatically manipulating the composition of the dark state. At the initial moment, the control field is strong, satisfying $\Omega_c \gg \sqrt{N}\,\Omega_p$. In this limit, the dark state is almost purely a photonic state. An incoming signal photon thus prepares the system in the initial state $\ket{G}$. Subsequently, the control field is adiabatically ramped down to zero. According to the adiabatic theorem, the system remains in the dark state, which smoothly evolves to become a purely collective atomic excitation as $\Omega_c \to 0$. This coherently maps the photonic state onto the atomic spin-wave state, thereby storing the signal field:
\begin{equation}
    \ket{G} \xrightarrow{\text{Adiabatic evolution as } \Omega_c \to 0} \ket{R}.
\end{equation}
After the desired storage time, the control field $\Omega_c$ is turned on to retrieve the stored signal.

In contrast, the bright states $\ket{\Psi_{\pm}(t)}$ possess a finite population in the excited state $\ket{E}$ and are therefore coupled to radiative decay, rendering them intrinsically lossy. Together, the set $\{ \ket{\Psi_0}, \ket{\Psi_+}, \ket{\Psi_-} \}$ forms a complete orthonormal basis of dressed states for the coupled light-matter system.

\subsection{Counterdiabatic Driving for Accelerated Rydberg-EIT Storage}
For fast control protocols, the ideal adiabatic evolution of the dark state is violated. In particular, a finite rate of change of the mixing angle
\begin{equation}
    \tan\theta(t) = \frac{\sqrt{N}\,\Omega_p(t)}{\Omega_c(t)},
\end{equation}
induces non-adiabatic couplings that drive the system out of the dark-state manifold, leading to an unwanted population of the lossy intermediate state $\ket{E}$. To suppress non-adiabatic transitions, we introduce a CD driving field defined as~\cite{Zhao2017Robust,RevModPhys.91.045001,Sun2020}
\begin{equation} \Omega_{\mathrm{CD}}(t) = \frac{\Omega_c(t)\,[\sqrt{N}\,\dot{\Omega}_p(t)] - [\sqrt{N}\,\Omega_p(t)]\,\dot{\Omega}_c(t)} {N\Omega_p^2(t) + \Omega_c^2(t)}. \end{equation}
The corresponding auxiliary Hamiltonian reads
\begin{equation} H_{\mathrm{CD}}(t) = i\,\frac{\Omega_{\mathrm{CD}}(t)}{2} \bigl(\ket{G}\!\bra{R}-\ket{R}\!\bra{G}\bigr), \end{equation}
describing an effective direct coupling between the ground state $\ket{G}$ and the collective Rydberg state $\ket{R}$. The detailed derivation of the CD driving field and the resulting auxiliary Hamiltonian is provided in the Appendix~\ref{App:A}. This additional control term exactly cancels the non-adiabatic couplings, thereby enforcing transitionless evolution along the instantaneous dark state, even in the non-adiabatic fast driving regime. Such CD driving is particularly well suited for the level structure of Rydberg atoms. To experimentally implement CD driving between the ground state $\ket{G}$ and the Rydberg state $\ket{R}$, we employ an effective two-photon transition with a large intermediate detuning $\Delta$, as illustrated in the level diagram of Fig.~\hyperref[fig1]{1(b)}. The corresponding pulse sequence for the writing and retrieval of the signal photon is schematically depicted in Fig.~\hyperref[fig1]{1(c)}. The explicit forms of the two-photon pulses, including those used for implementing the CD driving, are provided in Appendix~\ref{App:D}. In this configuration, the direct $\ket{G}\leftrightarrow\ket{R}$ coupling is realized via off-resonant excitation through the intermediate state while avoiding real population transfer to $\ket{E}$. In our ladder-type system configuration, the effective target wave vector for the spin wave is determined by the sum of the probe and coupling wave vectors $\mathbf{k}_{\text{eff}} = \mathbf{k}_p + \mathbf{k}_c$. Assuming a counter-linear propagating configuration, we can focus on the scalar magnitude of the wave vector, denoted as $k_{\text{eff}}$. However, the introduction of CD driving unavoidably contributes an additional wave vector, leading to a small shift in the effective spin-wave momentum. Using the ARC (Alkali.ne Rydberg Calculator) for $^{87}\text{Rb}$~\cite{ROBERTSON2021107814}, we calculated the corresponding optical wave vectors for excitation to the target Rydberg state $\ket{60S_{1/2}}$ by comparing the wave vectors for the ideal resonant path ($\Delta=0$) and the actual CD path with an intermediate detuning $\Delta/2 \pi = 10$ GHz. For the standard resonant condition, the wavelengths are $\lambda_p \approx 780.2415$ nm and $\lambda_c \approx 479.8389$ nm, yielding an effective wave vector $k_{\text{eff}} \approx 5,041,491.7$ rad/m. For the CD driving path ($\Delta/2 \pi = 10$ GHz), the shifted wavelengths are $\lambda'_p \approx 780.1139$ nm and $\lambda'_c \approx 479.8871$ nm, resulting in a modified wave vector $k_{\mathrm{CD}} \approx 5,038,858$ rad/m. The resulting wave vector mismatch $\Delta k$ is defined and calculated as follows:
\begin{equation}
    \Delta k = |k_{\text{eff}} - k_{\mathrm{CD}}| \approx 2633 \text{ rad/m}.
\end{equation}
Considering the length of the atomic medium in our simulations ($L=100\, \mu\text{m}$), the accumulated phase error is $\Delta \phi = \Delta k \cdot L$, which is negligible. Nevertheless, $\Delta k$ is retained in the simulations to explicitly verify that its effect on the spin-wave evolution remains minor.

\section{Results and Discussion}\label{sec3}

\subsection{Temporal Evolution of Atomic Populations}

\begin{figure}
	\centering
	\includegraphics[width=1\linewidth]{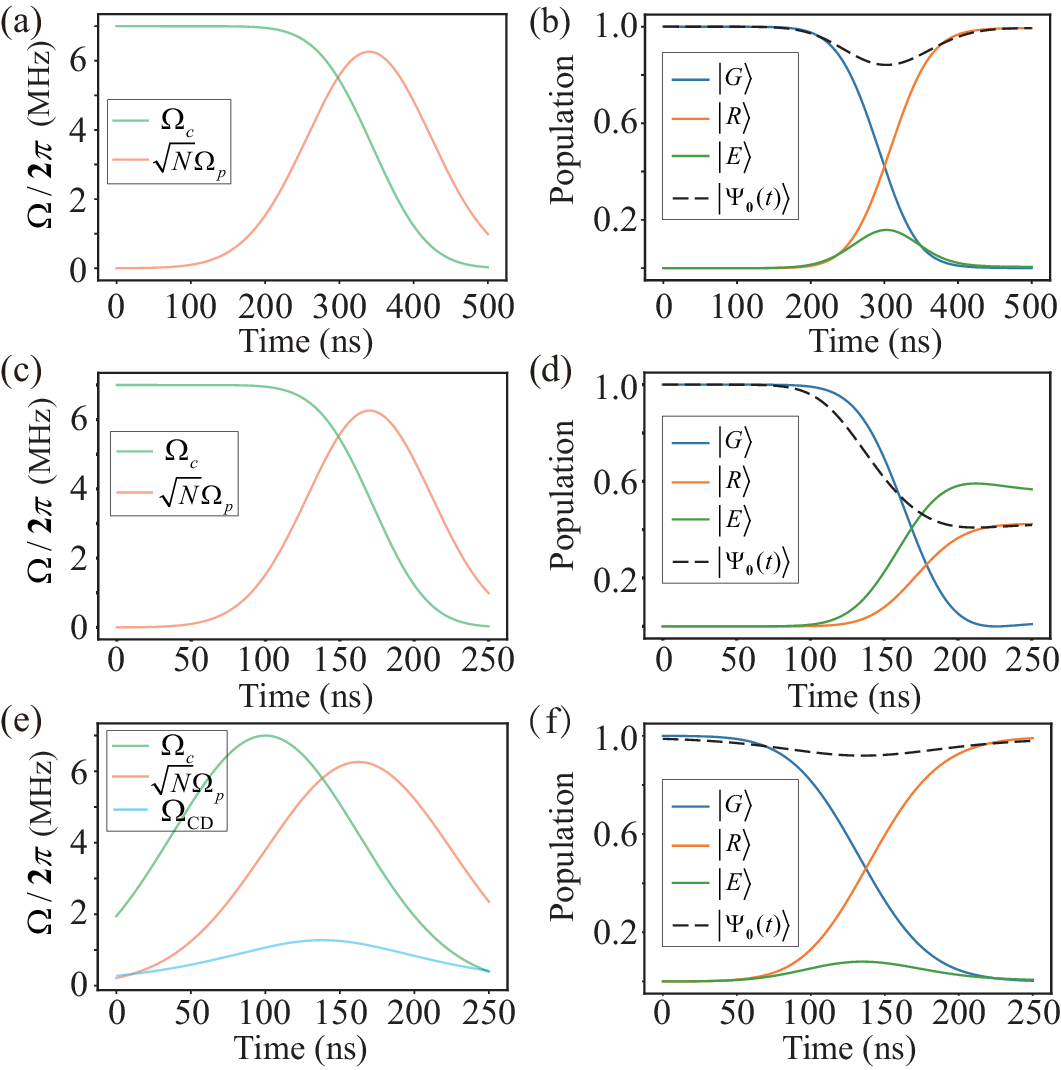}
	\caption{Population dynamics under different scenarios. (a) Standard pulse sequence with total duration $T=500~\mathrm{ns}$, consisting of a Gaussian probe field and a smoothed rectangular control field. (b) Population dynamics corresponding to the pulse sequence in (a). (c) The same pulse sequence temporally compressed to $T=250~\mathrm{ns}$. (d) Population dynamics corresponding to the pulse sequence in (c), where a clear reduction in transfer fidelity is observed due to the breakdown of adiabatic following. (e) CD pulse sequence at $T=250~\mathrm{ns}$, in which an auxiliary CD drive $\Omega_{\mathrm{CD}}$ (blue curve) is applied in addition to the base pulses. (f) Population dynamics under CD driving, demonstrating the restoration of high-fidelity population transfer. The black dashed curves indicate the instantaneous dark-state fidelity. The parameters used in the simulations are $\Omega_p/2 \pi=0.28$ MHz and  $\Omega_c/2 \pi=7$ MHz.}\label{fig2}
\end{figure}

To explore the dynamics of Rydberg excitation and the limitations of adiabatic population transfer, we first analyze the system evolution under a standard experimental pulse sequence. Figure~\hyperref[fig2]{2(a)} shows the time-dependent Rabi frequencies for a total operation time of $T=500~\mathrm{ns}$. The probe field $\Omega_p(t)$ (red curve) is modeled as a Gaussian pulse centered at $t_p = 0.68T$ with a standard deviation $\sigma_p = T/4$. To mimic realistic experimental conditions in which the control field is switched on before the probe and turned off adiabatically, the control field $\Omega_c(t)$ (green curve) is described by a smoothed rectangular pulse constructed using error functions~\cite{pulse_parameters_2026}. The corresponding population dynamics for $T=500~\mathrm{ns}$ is shown in Fig.~\hyperref[fig2]{2(b)}. In this pulse sequence, the system undergoes a smooth adiabatic evolution from the ground state $|G\rangle$ to the target Rydberg state $|R\rangle$. The adiabatic condition is satisfied, leading to a strong suppression of the intermediate-state population $P_E$ (green curve). Consequently, the final Rydberg state population $P_R$ (orange curve) exceeds $99\%$, and the fidelity with respect to the instantaneous dark state remains close to unity throughout the evolution. The dynamics of the system change significantly when the total operation time is reduced to $T=250~\mathrm{ns}$ to accelerate the protocol, as shown in Fig.~\hyperref[fig2]{2(c)}. In this fast evolution regime, the increased time derivative of the mixing angle gives rise to significant non-adiabatic couplings, causing the system to deviate from the dark-state manifold. As shown in Fig.~\hyperref[fig2]{2(d)}, the final population transferred to $|R\rangle$ drops to approximately $50\%$, accompanied by substantial leakage into the lossy intermediate state $|E\rangle$ and a rapid decline in the fidelity of the dark state. These results clearly demonstrate the breakdown of adiabatic following for short operation times.

To restore high-fidelity population transfer in the non-adiabatic regime, we implement CD driving. The CD pulse sequence for $T=250~\mathrm{ns}$ is shown in Fig.~\hyperref[fig2]{2(e)}, where the base pulses are considered to be two partially overlapping Gaussian envelopes to facilitate the analytic construction of the CD field. Specifically, the probe and control pulses are centered at $t_p = 0.65T$ and $t_c = 0.40T$, respectively, with the control pulse having a standard deviation $\sigma_c = T/4$. The resulting population dynamics are presented in Fig.~\hyperref[fig2]{2(f)}. With CD driving, leakage to the intermediate state is strongly suppressed, and the final Rydberg state population $P_R$ reaches approximately $99\%$. Importantly, the fidelity with respect to the instantaneous dark state remains close to unity throughout the evolution, demonstrating that the CD field effectively compensates for the non-adiabatic couplings induced by the rapid temporal modulation.

\subsection{Maxwell-Bloch Propagation Analysis}

Heretofore, our analysis has focused on the population dynamics of the atomic states, with a particular emphasis on the effects of rapid driving and STA compensation. In contrast to protocols where the primary concern is the redistribution of populations, such as in STIRAP, the objective of photon storage is fundamentally different: it aims to coherently map an incoming signal field onto a collective atomic spin wave. For this purpose, the temporal and spatial characteristics of the signal field itself become critical, as they directly determine the performance of the storage process. This distinction highlights a fundamental difference between EIT-based storage schemes and STIRAP, which primarily governs population transfer. In the following section, we provide a detailed analysis of signal field propagation and its storage within the Rydberg superatoms.

Considering the collective nature of the Rydberg blockade, we adopt the superatom model, in which the coupling between the probe field and the atomic ensemble is collectively enhanced by a factor of $\sqrt{N}$. Under the two-photon resonance condition, the atomic dynamics are described by a set of modified Optical Bloch Equations (OBEs). In addition to the stored probe field, the CD field contributes to effective ground-Rydberg coupling. The spatial phase factor $e^{\pm i\Delta k z}$ arising from the wave-vector mismatch associated with the CD driving path, as discussed in the preceding section, is also incorporated into the atomic coherences. With these considerations, the density-matrix elements $\rho_{ij}$ obey the following OBEs:
\begin{align}
\frac{\partial}{\partial t}\rho_{rg} =& \frac{i}{2}\Omega_c \rho_{eg} -  \frac{\Gamma_r}{2}\rho_{rg} -  \frac{i}{2} \sqrt{N} \Omega_p \rho_{re}\nonumber \\
&- \frac{1}{2}\Omega_{\mathrm{CD}} e^{i \Delta k z}(\rho_{gg} - \rho_{rr}), \nonumber \\
\frac{\partial}{\partial t}\rho_{eg} =& \frac{i}{2}\sqrt{N}\Omega_p (\rho_{gg} - \rho_{ee}) + \frac{i}{2}\Omega_c \rho_{rg} - \frac{\Gamma}{2}\rho_{eg} \nonumber \\
&+ \frac{1}{2}\Omega_{\mathrm{CD}} e^{i \Delta k z}\rho_{er}, \nonumber \\
\frac{\partial}{\partial t}\rho_{er} =& \frac{i}{2}\sqrt{N}\Omega_p \rho_{rg}^* + \frac{i}{2}\Omega_c (\rho_{rr} - \rho_{ee}) - \frac{\Gamma+\Gamma_r}{2}\rho_{er} \nonumber \\
&- \frac{1}{2}\Omega_{\mathrm{CD}} e^{-i \Delta k z}\rho_{eg}, \nonumber \\
\frac{\partial}{\partial t}\rho_{rr} =& \frac{i}{2}\Omega_c \rho_{er} - \frac{i}{2}\Omega_c^* \rho_{er}^* - \Gamma_r \rho_{rr} \nonumber \\
&-\frac{1}{2} \Omega_{\mathrm{CD}} (\rho_{rg} e^{-i \Delta k z}
+ \rho_{rg}^* e^{i \Delta k z}), \nonumber \\
\frac{\partial}{\partial t}\rho_{ee} =& -\frac{i}{2}\sqrt{N}\Omega_p^* \rho_{eg} + \frac{i}{2}\sqrt{N}\Omega_p \rho_{eg}^* \nonumber  \nonumber \\
&- \frac{i}{2}\Omega_c \rho_{er} + \frac{i}{2}\Omega_c^* \rho_{er}^* - \Gamma \rho_{ee}, \nonumber \\
\frac{\partial}{\partial t}\rho_{gg} =& \frac{i}{2}\sqrt{N}\Omega_p^* \rho_{eg} - \frac{i}{2}\sqrt{N}\Omega_p \rho_{eg}^* + \Gamma \rho_{ee}+\Gamma_r\rho_{rr}\nonumber \\
&+\frac{1}{2} \Omega_{\mathrm{CD}} (\rho_{rg} e^{-i \Delta k z}
+ \rho_{rg}^* e^{i \Delta k z}).
\end{align}
Here, $\Gamma$ and $\Gamma_r$ denote the decay rates of the intermediate state and the Rydberg state, respectively. The detailed derivation of these equations, including the incorporation of the CD-induced phase factors, is provided in Appendix~\ref{App:B}. The propagation of the probe field $\Omega_p$ through the medium is described by the Maxwell-Schr\"odinger equation (MSE):
\begin{equation}
\left( \partial_z + \frac{1}{c}\partial_t \right)\Omega_p=i \frac{\alpha \Gamma }{2\sqrt{N} L}\,\rho_{eg},
\label{eq:MSE}
\end{equation}
where $c$ is the speed of light in vacuum, $\alpha$ is the optical depth, and $L$ is the length of the medium. The above propagation equation is expressed in the collective basis, where the light–matter coupling is enhanced by $\sqrt{N}$. 
A detailed derivation of the MSE in this collective framework is given in the Appendix~\ref{App:C}. The DSP field operator can be expressed as $\Psi(z,t) = \cos\theta(z,t)\,\Omega_p(z,t) - \sin\theta(z,t)\,\sqrt{N}\,\rho_{rg}(z,t)$ with $\theta(z,t)={\arctan}[\sqrt{N}\Omega_p(z,t)/\Omega_c(z,t)]$. At the initial time, in the limit of a strong control field $\Omega_c \gg \sqrt{N}\Omega_p$, the polariton is predominantly photonic and propagates through the medium with a finite group velocity. Conversely, when the control field is adiabatically switched off, the polariton is completely mapped to the collective Rydberg coherence $\rho_{rg}$, becoming stationary with a vanishing group velocity $v_g \to 0$. The subsequent reapplication of the control field enables coherent retrieval of the stored excitation, realizing a reversible light-matter interface.

\begin{figure}
	\centering
	\includegraphics[width=1\linewidth]{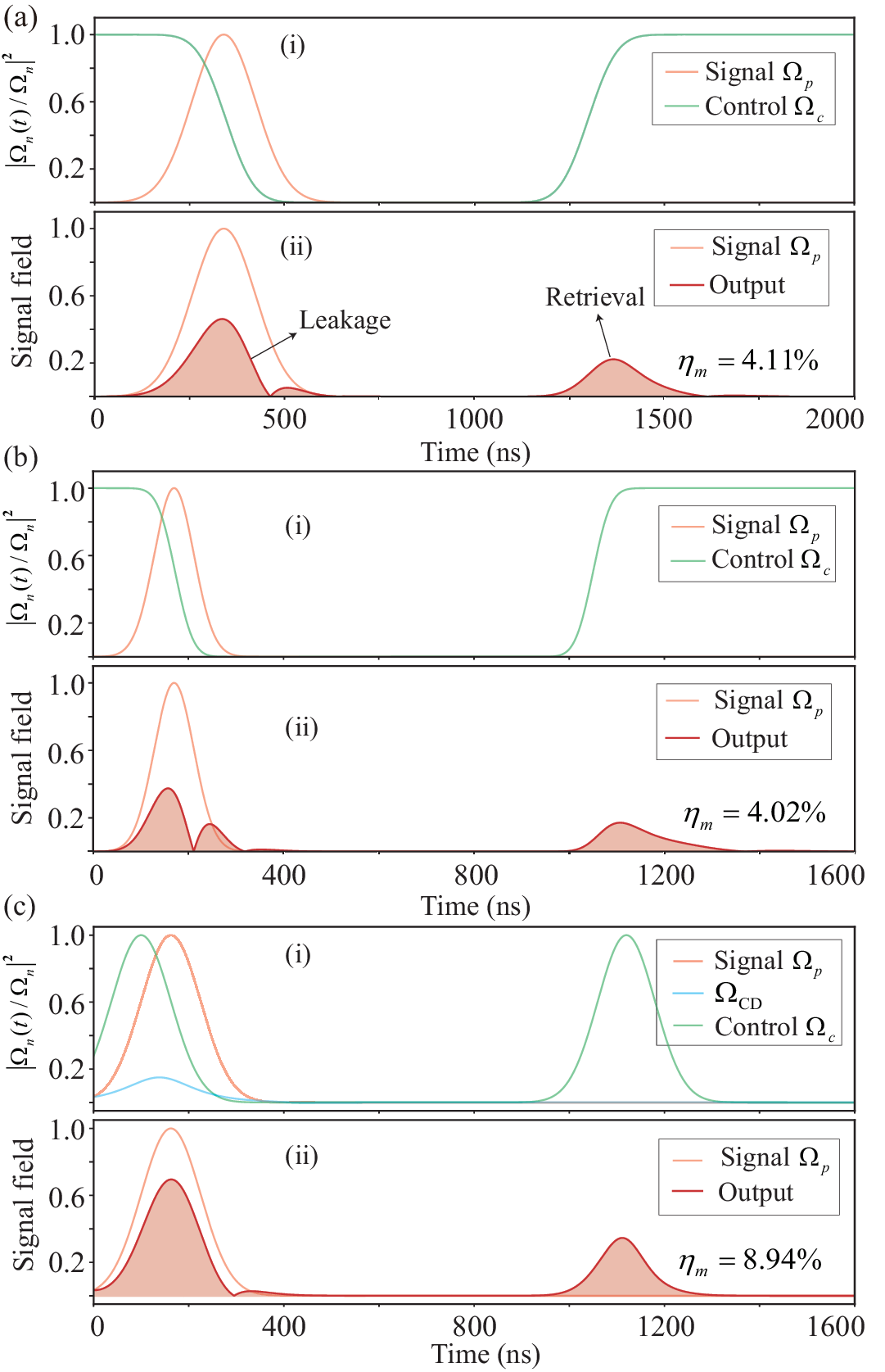}
	\caption{Comparison of EIT light storage protocols. (a) Conventional scheme with a writing time of 500 ns: (i) Pulse sequence of the input signal field and the control field; (ii) The retrieved output signal field. (b) Conventional scheme with a shorter writing time of 250 ns: (i) Pulse sequence of the input signal field and the control field; (ii) Retrieved output signal field. (c) CD-driven scheme: (i) Pulse sequence with CD driving; (ii) The corresponding retrieved signal field. The parameters used in the simulations are $\Gamma_r=2\pi \times 10$ kHz, $\Gamma=2\pi \times 6$ MHz, $\alpha=5$, $N=500$, and $L=100\,\mu\mathrm{m}$.}\label{fig3}
\end{figure}

Figure~\hyperref[fig3]{3(a)} illustrates the pulse sequence and signal-field dynamics under the conventional EIT-based optical storage protocol with a writing time of $T=500$~ns. As shown in Fig.~\hyperref[fig3]{3(a-i)}, an adiabatic pulse-control strategy is employed: during the writing stage, the signal field enters the atomic medium while the control field is initially turned on. As the control field gradually decreases, the photonic excitation is coherently mapped onto the collective Rydberg spin-wave of the atomic ensemble, which corresponds to the storage of the signal photons. By solving the Maxwell--Schr\"{o}dinger equations, we obtain the temporal evolution of the signal field, shown in Fig.~\hyperref[fig3]{3(a-ii)}. During the writing process (at approximately $T_{\mathrm{writing}} = 500$~ns), a fraction of the signal field is not fully converted into spin-wave excitations owing to the finite optical depth and residual non-adiabatic effects, resulting in direct leakage through the medium. After a controlled storage interval, the control field is switched on again during the retrieval stage, and the stored spin-wave excitation is converted back into a propagating optical field. Due to the phase-matching condition, the retrieved signal is emitted along the original propagation direction. {We evaluate the storage performance using the quantum memory storage efficiency defined as~\cite{Reim2010,PhysRevLett.107.053603,Wang2019}
\begin{equation}
\eta_m = \frac{\int |\psi_{\text{out}}(\tau)|^2 d\tau}{\int |\psi_{\text{in}}(\tau)|^2 d\tau},
\end{equation}
where $\psi_{\text{out}}(\tau)$ and $\psi_{\text{in}}(\tau)$ denote the temporal wavefunctions of the retrieved and input signal fields, respectively. Under the present conditions, the storage efficiency is approximately $4.11\%$.} To examine the effect of shortening the writing time, Fig.~\hyperref[fig3]{3(b)} shows the same conventional EIT protocol with a reduced writing time of $T_{\mathrm{writing}}=250$~ns. The pulse sequence is displayed in Fig.~\hyperref[fig3]{3(b-i)}, and the corresponding retrieved signal field is shown in Fig.~\hyperref[fig3]{3(b-ii)}. Compared with the 500~ns case, the retrieved signal becomes noticeably weaker when the pulse duration is shortened, {with the corresponding storage efficiency reduced to approximately $4.02\%$.} This reduction arises because rapid temporal variation violates the adiabatic condition, resulting in a significant transient population in the intermediate state. As a result, a considerable fraction of the signal excitation is lost through spontaneous scattering from the intermediate state, rather than being coherently mapped onto the collective atomic spin wave.

To accelerate the writing process while overcoming the adiabatic limitations of conventional EIT storage, we introduce a CD driving field, as shown in Fig.~\hyperref[fig3]{3(c)}. The modified pulse sequence is shown in Fig.~\hyperref[fig3]{3(c-i)}, where the blue curve denotes the auxiliary CD field. This field predominantly acts on the rapidly varying edges of the pulse sequence, compensating for non-adiabatic transitions and suppressing unwanted population leakage during fast evolution. The corresponding signal-field dynamics are shown in Fig.~\hyperref[fig3]{3(c-ii)}. In contrast to the conventional EIT protocol, the presence of a CD drive enables both storage and retrieval to be completed on a significantly shorter timescale. Although CD driving introduces additional coupling terms in the effective Hamiltonian and involves a correction to the effective wave-vector mismatch $\Delta k$, this correction does not alter the retrieval mechanism in our scheme. When the control field is switched on during retrieval, the stored spin wave continues to satisfy the phase-matching condition and is faithfully converted back into the original signal field mode, with a storage efficiency of $\eta_m = 8.94\%$ under the same experimental parameters and conditions as the conventional EIT storage scheme described above. As a result, even for significantly shortened operation times, the CD-assisted scheme yields a retrieved signal with much higher efficiency than that obtained in conventional EIT storage. This demonstrates that the CD field serves as an auxiliary tool that accelerates the dynamics and enhances the storage fidelity without introducing additional mode distortion or aliasing in the retrieved optical signal. For completeness, we discuss the numerical implementation of the effective CD field $\Omega_{\mathrm{CD}}$ using two Raman fields with large detunings, as detailed in the Appendix~\ref{App:D}.

\subsection{Signal Retrieval and Storage Performance}

\begin{figure}
	\centering
	\includegraphics[width=1\linewidth]{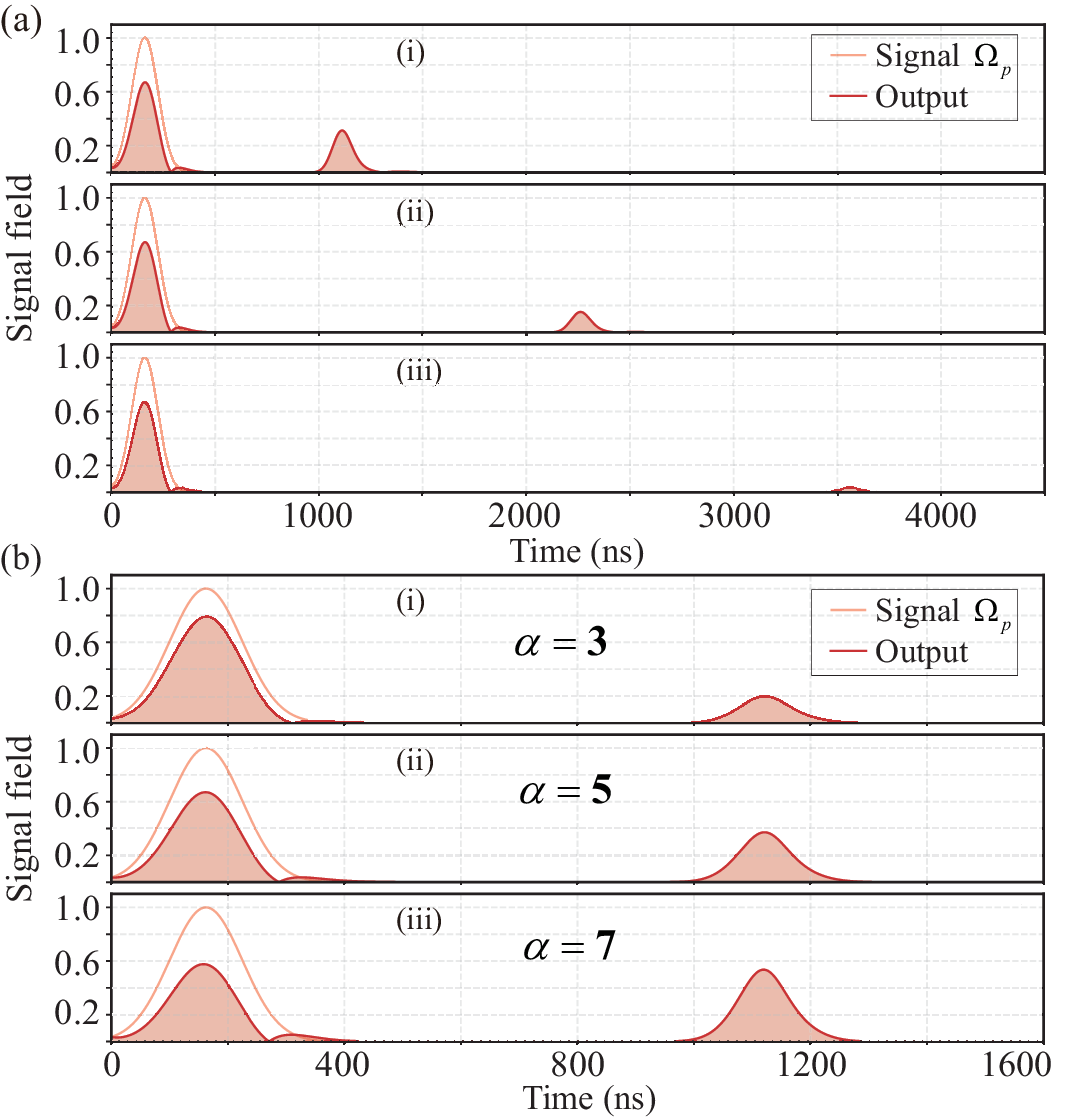}
	\caption{Retrieved output signal fields under the CD driving protocol with varying storage parameters.
(a) Output fields for different storage times at a fixed optical depth ($\alpha=5$). Panels (i), (ii), and (iii) correspond to storage times of $t_s=800$ ns, $1500$ ns, and $3000$ ns, respectively. (b) Output fields for different optical depths at a fixed storage time ($t_s=800$ ns). Panels (i), (ii), and (iii) correspond to optical depths of $\alpha=3, 5,$ and $7$, respectively. Other parameters are the same with Fig.~\ref{fig3}.}\label{fig4}
\end{figure}

To systematically evaluate the storage and retrieval performance of the proposed scheme, we investigate the temporal evolution of the retrieved signal field for different storage durations. In addition to the spontaneous decay of the Rydberg state, which has been considered in many previous studies, our analysis explicitly accounts for the effect of the collective motional dephasing of the Rydberg spin wave~\cite{PhysRevA.108.042405,PhysRevLett.134.053604}. After the signal photon is mapped onto the atomic ensemble, the subsequent thermal motion of the atoms leads to a gradual loss of phase coherence due to position-dependent phase accumulation. The creation operator of the Rydberg spin wave at the storage time $t_s$ can be written as
\begin{equation}
S^\dagger(t_s)
= \frac{1}{\sqrt{N}} \sum_{j=1}^{N}
e^{i k_{\mathrm{eff}} [z_j(0) + v_j t_s]}
\left( |r\rangle\langle g| \right)_j ,
\end{equation}
where $z_j(0)$ and $v_j$ denote the initial position and the velocity of the $j$th atom along the propagation direction, respectively, and $k_{\mathrm{eff}}$ is the effective wave vector of the Rydberg spin wave. For an ensemble of $^{87}\mathrm{Rb}$ atoms cooled to a typical temperature of $T_a = 10~\mu\mathrm{K}$, the root-mean-square velocity is $
v = \sqrt{{k_B T_a}/{m}} $, where $k_B$ is the Boltzmann constant and $m$ is the atomic mass. Averaging over the thermal velocity distribution yields a Gaussian decay of the collective coherence, $\langle S(t_s)S^\dagger(0) \rangle
\propto \exp\!\left[-\tfrac{1}{2}(k_{\mathrm{eff}} v)^2 t_s^2\right]$, which defines a characteristic motional dephasing time scale $T_2^* \sim 1/(k_{\mathrm{eff}} v)$. Since directly tracking the motion of a large number of individual atoms is computationally impractical, we adopt a macroscopic effective model. By averaging the random thermal motions, the inhomogeneous dynamics of the ensemble can be reduced to a collective Bloch equation. Consequently, in our numerical simulations, this collective motional dephasing is incorporated into the optical Bloch equations as a time-dependent dephasing term acting on the ground-to-Rydberg coherence $\rho_{rg}$. Explicitly, its equation of motion reads
\begin{align}
\frac{\partial}{\partial t}\rho_{rg}
= &\frac{i}{2} \Omega_c \rho_{eg}
- \frac{1}{2}\Omega_{\mathrm{CD}} e^{i \Delta k z} (\rho_{gg} - \rho_{rr}) \nonumber \\
&- \left[ \frac{\Gamma_r}{2} + (k_{\mathrm{eff}} v)^2 t \right] \rho_{rg}-  \frac{i}{2} \sqrt{N} \Omega_p \rho_{re}.
\end{align}
The explicitly time-dependent term $(k_{\mathrm{eff}} v)^2 t$ corresponds to the instantaneous dephasing rate that reproduces the Gaussian decay of the collective coherence. As a direct consequence of this motional dephasing, the retrieval efficiency decreases rapidly with increasing storage time, as observed in Figs.~\hyperref[fig4]{4(a-i)}–\hyperref[fig4]{4(a-iii)}. For a storage duration of $t_s = 3000~\mathrm{ns}$, the accumulated phase diffusion is sufficiently strong that the collective phase information of the spin wave is almost completely erased, resulting in a nearly vanishing retrieval signal. This behavior is consistent with typical experimental observations, where spin-wave dephasing caused by finite temperature and atomic motion severely limits the achievable storage lifetime under similar conditions~\cite{PhysRevX.5.031015}.

In addition to the storage time, the storage efficiency is also strongly influenced by the optical depth of the atomic ensemble. A larger optical depth generally enhances the light–matter interaction and improves the efficiency of mapping the signal field to the Rydberg collective spin wave. In practice, achieving a higher effective optical depth of the superatom would therefore be beneficial for the storage performance. To examine this effect within our scheme, we further analyze the dependence of the signal retrieval on the optical depth, as shown in Fig.~\hyperref[fig4]{4(b)}. In this simulation, the storage time is fixed at $t_s = 800$~ns, while the optical depth is varied among $\alpha=3$, $5$, and $7$. The retrieved signal fields shown in Figs.~\hyperref[fig4]{4(b-i)}–\hyperref[fig4]{4(b-iii)} clearly demonstrate that increasing the optical depth results in a pronounced enhancement of the retrieval signal intensity. This improvement can be attributed to the stronger light–matter coupling provided by a larger optical depth, which suppresses signal leakage during the writing stage and thereby enhances both storage and retrieval efficiencies. It is worth emphasizing that the optical depth range considered in Fig.~\hyperref[fig4]{4(b)} is chosen to realistically reflect experimentally accessible conditions. In superatom-based Rydberg systems, the effective optical depth is typically limited to values below $\alpha\sim 10$, due to constraints imposed by the Rydberg blockade radius and achievable atomic densities~\cite{Yang:22,Kumlin_2023,10.1063/5.0211071,cmpn-jfqr,2sl5-7pbr}. Our simulations demonstrate that even in an experimentally relevant and relatively modest optical depth regime, the introduction of CD driving enables efficient and robust signal storage and retrieval.

\begin{figure}
	\centering
	\includegraphics[width=1\linewidth]{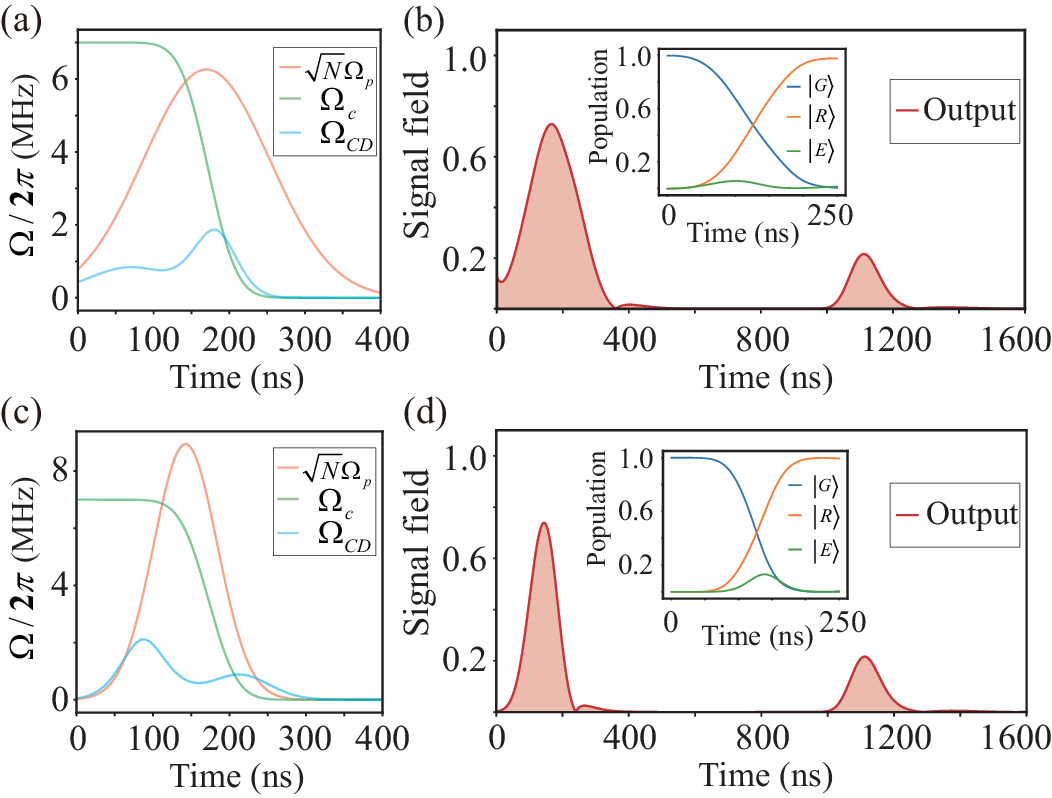}
	\caption{(a,c) Applied pulse sequences composed of a smoothed square control field, a CD drive, and two different Gaussian probe pulses. Panels (a) and (c) correspond to the two probe pulses with different temporal profiles. (b,d) Corresponding output probe fields obtained by solving the propagation equations for the pulse sequences shown in (a) and (c), respectively. The insets display the population dynamics of the states $\ket{G}$, $\ket{E}$, and $\ket{R}$. In addition to pulse parameters, other parameters are the same with Fig.~\ref{fig3}.}\label{fig5}
\end{figure}

To verify the universality and robustness of the proposed scheme with respect to different pulse waveforms, we replace the control field with a smoothed square pulse and investigate the system response to probe fields with different temporal envelopes. Figures~\hyperref[fig5]{5(a)} and ~\hyperref[fig5]{5(b)} present the first scenario. Here, the input probe pulse has a relatively broad temporal profile with a standard deviation $\sigma_p = T_{\mathrm{write}}/3$ and a maximum Rabi frequency of $2\pi \times 0.28~\mathrm{MHz}$. The corresponding CD driving field, calculated from the probe and control pulses, is shown as the blue curve in Fig.~\hyperref[fig5]{5(a)}. Figure~\hyperref[fig5]{5(b)} shows the simulated retrieval signal. Under this pulse configuration, the signal field can still be written into the atomic ensemble in a short time and efficiently retrieved during the readout process. This indicates that the proposed scheme maintains robust storage and retrieval performance under such input conditions. The inset of Fig.~\hyperref[fig5]{5(b)} shows the corresponding population dynamics of the atomic states during the writing stage. Within a writing window of approximately $250~\mathrm{ns}$, the system closely follows the instantaneous dark state, while the population of the lossy intermediate state $\ket{E}$ remains strongly suppressed. Consequently, the population of the Rydberg state $\ket{R}$ at the end of the writing process, which corresponds to the storage fidelity, approaches unity. To evaluate the performance of the scheme under a different probe field configuration, we consider a narrower input signal pulse with a standard deviation $\sigma_p = T_{\mathrm{write}}/6$ and increase its maximum Rabi frequency to $ 2\pi \times 0.4~\mathrm{MHz}$, as shown in Fig.~\hyperref[fig5]{5(c)}. The corresponding CD driving field, calculated from the probe and control pulses, is shown as the blue curve in Fig.~\hyperref[fig5]{5(c)}. The resulting retrieval signal in Fig.~\hyperref[fig5]{5(d)} shows that the signal field can still be rapidly written into the atomic ensemble and efficiently retrieved. The inset further displays the population dynamics, where the system closely follows the instantaneous dark state and the population of the lossy intermediate state remains strongly suppressed, leading to high-fidelity population transfer to the Rydberg state at the end of the writing stage. In general, these results indicate that the proposed CD driving scheme is not restricted to a specific pulse configuration. Even for probe pulses with different temporal widths, the introduction of the CD drive enables fast and high-fidelity optical storage and retrieval.

\subsection{Robustness against CD Control Errors}

\begin{figure}
	\centering
	\includegraphics[width=1\linewidth]{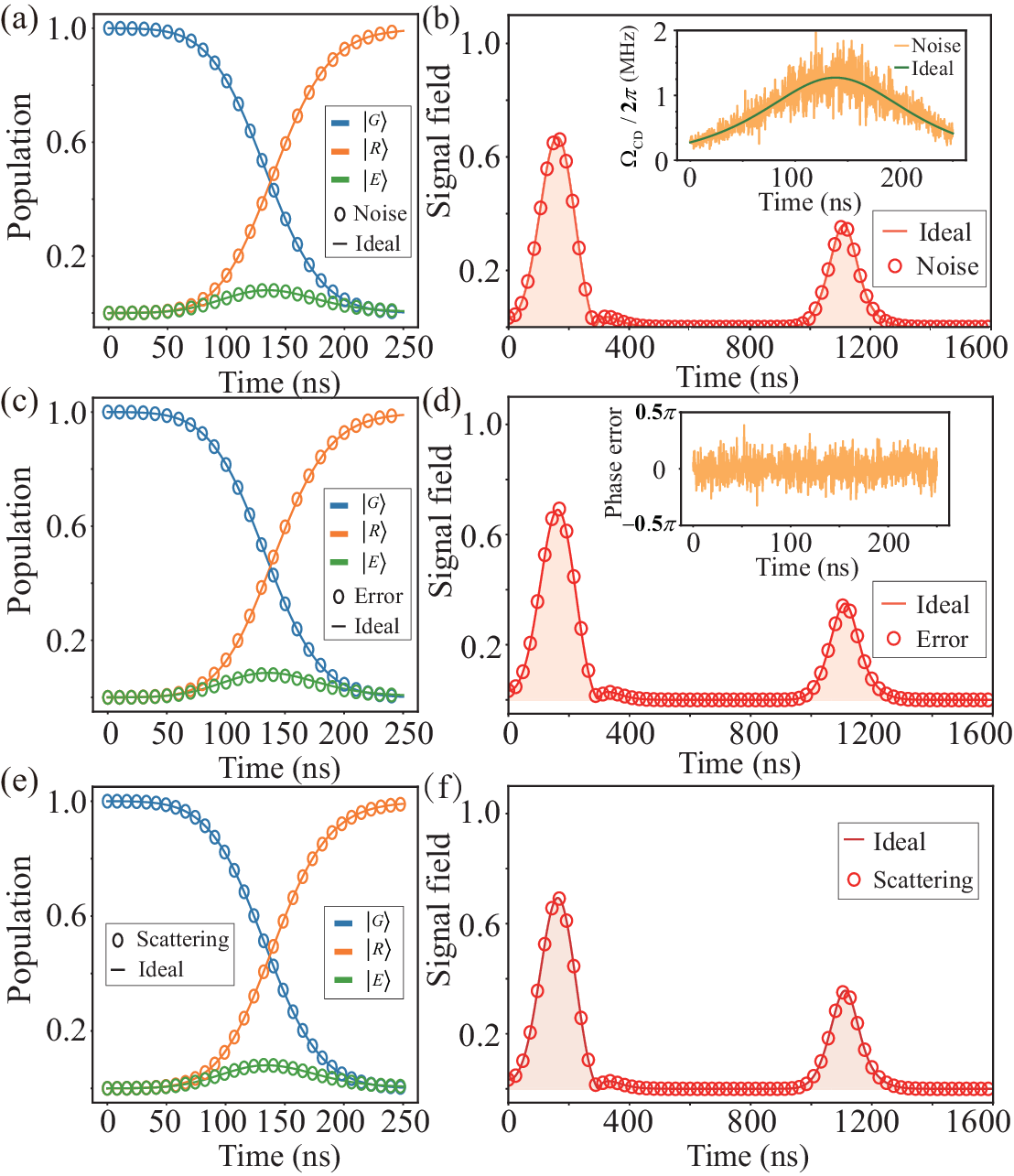}
	\caption{Robustness analysis of the STA-assisted storage protocol under imperfections in the CD driving field. (a) Population dynamics with Rabi-amplitude noise in the CD field. (b) Retrieved signal obtained from the propagation equation under the same amplitude noise, with the inset showing the CD field under Rabi-amplitude fluctuations. (c) Population dynamics with a CD phase error. (d) Retrieved signal corresponding to the phase errors, with the inset showing the CD field under phase errors. (e) Population dynamics and (f) retrieved signal in the presence of off-resonant scattering from the intermediate state. Other parameters are the same with Fig.~\ref{fig3}.}\label{fig6}
\end{figure}

The above results are obtained under the assumption of an ideal CD field. In practice, however, the exact atom number in the ensemble is not perfectly known, and the driving lasers are subject to experimental imperfections, which lead to deviations of the implemented CD field from its ideal form. To evaluate the robustness of the proposed scheme under such conditions, we numerically study the effects of Rabi amplitude noise, phase errors, and off-resonant scattering on the CD field. Figures~\hyperref[fig6]{6(a)} and~\hyperref[fig6]{6(b)} show the results obtained in the presence of Rabi-amplitude fluctuations in the CD driving field. In the simulation, the CD field is modified as $\Omega_{\mathrm{CD}}\rightarrow (1+\eta)\Omega_{\mathrm{CD}}$, where $\eta$ represents a random fluctuation within $\pm20\%$. The results are obtained by averaging over 30 independent realizations of the noise. As illustrated in Fig.~\hyperref[fig6]{6(a)}, the population transfer to the Rydberg state remains high-fidelity, reaching a final value of approximately $99\%$ despite the substantial amplitude noise. The corresponding retrieved signal, calculated from the propagation equation and displayed in Fig.~\hyperref[fig6]{6(b)}, shows almost no noticeable distortion compared with the ideal case. We further examine the effect of phase errors in the CD driving field by introducing a fluctuating laser phase $\phi(t)=\phi_0+\delta\phi(t)$, which results in a modified Rabi frequency $\Omega_{\mathrm{CD}}\rightarrow \Omega_{\mathrm{CD}}e^{i\delta\phi(t)}$. The phase fluctuation is modeled as a zero-mean Gaussian random variable with a standard deviation of $\sigma = 0.1\pi$, which can be mathematically denoted as $\mathcal{N}(0, \sigma^2)$, and the results shown in Figs.~\hyperref[fig6]{6(c)} and~\hyperref[fig6]{6(d)} are obtained by averaging over 30 independent realizations. As shown in Fig.~\hyperref[fig6]{6(c)}, the population transfer to the Rydberg state remains of high fidelity, reaching a final value of about $99\%$. The corresponding retrieved signal in Fig.~\hyperref[fig6]{6(d)} exhibits only a negligible deviation from the ideal case. These results indicate that the STA-assisted storage protocol is robust against both Rabi-amplitude fluctuations and phase errors in the CD driving field.

In a realistic two-photon implementation of the CD field, the detrimental effect of off-resonant scattering from the intermediate state should be taken into account. Although the auxiliary fields operate in a far-detuned regime, they can still induce a small transient population in the intermediate state, which gives rise to residual off-resonant scattering. For an off-resonant two-photon process, the corresponding intermediate-state scattering rate can be estimated as~\cite{RevModPhys.82.2313,evered2023high,7zjs-73qm}
\begin{equation}
\gamma_{\text{IS}}(t)
=
\frac{|\Omega_p'(t)|^2+|\Omega_c'(t)|^2}{4\Delta^2}\Gamma ,
\end{equation}
where $\Omega_p'(t)$ and $\Omega_c'(t)$ are the single-photon Rabi frequencies of the auxiliary fields used to realize the effective CD coupling, $\Delta$ is the single-photon detuning, and $\Gamma$ is the natural linewidth of the intermediate state. From the perspective of collective excitation, such scattering events introduce random local phase perturbations to individual atoms and therefore degrade the phase coherence of the spin wave. In our theoretical model, this effect is consequently incorporated as an effective pure-dephasing process acting on the Rydberg-state coherence, described by the Lindblad term $\gamma_{\text{IS}}(t)\mathcal{D}[|r\rangle\langle r|]\rho$ in the master equation. As shown in Fig.~\hyperref[fig6]{6(e)} for the population dynamics and Fig.~\hyperref[fig6]{6(f)} for the signal propagation, the deviation between the ideal evolution and the scattering-included evolution is nearly imperceptible. This robustness can be attributed to two main factors. First, the single-photon detuning is chosen to be very large, $\Delta/2\pi=10~\mathrm{GHz}$, which strongly suppresses the intermediate-state scattering rate through the $1/\Delta^2$ scaling. Second, the CD field acts only as an auxiliary control field that compensates for nonadiabatic transitions and therefore requires relatively weak driving amplitudes. These results show that the proposed protocol remains robust not only against Rabi-amplitude noise and phase errors, but also against off-resonant intermediate-state scattering associated with the CD driving field.

\subsection{Performance under Multiphoton Excitations and Imperfect Rydberg Blockade}

In this section, we investigate the detrimental effects of an imperfect Rydberg blockade and a non-ideal single-photon input on the overall storage performance. When the incident signal is prepared as a weak coherent state rather than a pure single photon, its inherent Poissonian statistics introduce a non-vanishing probability of containing multiple photons, which can lead to a nonzero probability of multiple excitations. To account for such multiphoton components within our semi-classical framework, we introduce a weak-field amplitude correction factor $\epsilon$. Additionally, when the Rydberg blockade is imperfect and the van der Waals interaction $U_{rr}$ is not sufficiently large, a strong control field can further couple the system to states with double Rydberg excitations. This breaks the ideal adiabatic dark-state evolution and significantly increases the probability of double-excitation leakage.

To explicitly model these mechanisms, we expand the conventional single-excitation subspace $\{|G\rangle, |E\rangle, |R\rangle\}$ to include the double-excitation manifold. Assuming a symmetric uniform coupling, the newly added collective double-excitation states are defined as:
\begin{align}
    |EE\rangle &= \sqrt{\frac{2}{N(N-1)}} \sum_{i<j} |g_1 \dots e_i \dots e_j \dots g_N\rangle, \nonumber \\
    |RE\rangle &= \frac{1}{\sqrt{N(N-1)}} \sum_{i \neq j} |g_1 \dots r_i \dots e_j \dots g_N\rangle, \nonumber \\
    |RR\rangle &= \sqrt{\frac{2}{N(N-1)}} \sum_{i<j} |g_1 \dots r_i \dots r_j \dots g_N\rangle,
\end{align}
where $|EE\rangle$ represents the double intermediate state, $|RE\rangle$ is the mixed double-excitation state, and $|RR\rangle$ denotes the double Rydberg excitations. In the expanded $6 \times 6$ collective basis $\{|G\rangle, |E\rangle, |R\rangle, |EE\rangle, |RE\rangle, |RR\rangle\}$, the effective Hamiltonian in the rotating frame is given by:
\begin{widetext}
\begin{align}
H_T = -\frac{1}{2}
\begin{pmatrix}
0 & \sqrt{N}\Omega_p & -i\Omega_{\mathrm{CD}} & 0 & 0 & 0 \\
\sqrt{N}\Omega_p^* & 0 & \Omega_c & \epsilon \sqrt{2(N-1)}\Omega_p & 0 & 0 \\
i\Omega_{\mathrm{CD}} & \Omega_c^* & 0 & 0 & \epsilon \sqrt{N-1}\Omega_p & 0 \\
0 & \epsilon \sqrt{2(N-1)}\Omega_p^* & 0 & 0 & \sqrt{2}\Omega_c & 0 \\
0 & 0 & \epsilon \sqrt{N-1}\Omega_p^* & \sqrt{2}\Omega_c^* & 0 & \sqrt{2}\Omega_c \\
0 & 0 & 0 & 0 & \sqrt{2}\Omega_c^* & -2U_{rr}
\end{pmatrix}.
\end{align}
\end{widetext}
In this generalized Hamiltonian, the correction factor $\epsilon$ scales the transitions from the single- to the double-excitation manifold, thus introducing photon-number-dependent leakage channels. The diagonal term $2U_{rr}$ represents the finite Rydberg interaction energy. When $U_{rr}$ is not large enough to ensure a perfect blockade, the population can leak into double Rydberg excitation states, leading to additional spontaneous-emission loss and degradation of the signal.

\begin{figure}
	\centering
	\includegraphics[width=1\linewidth]{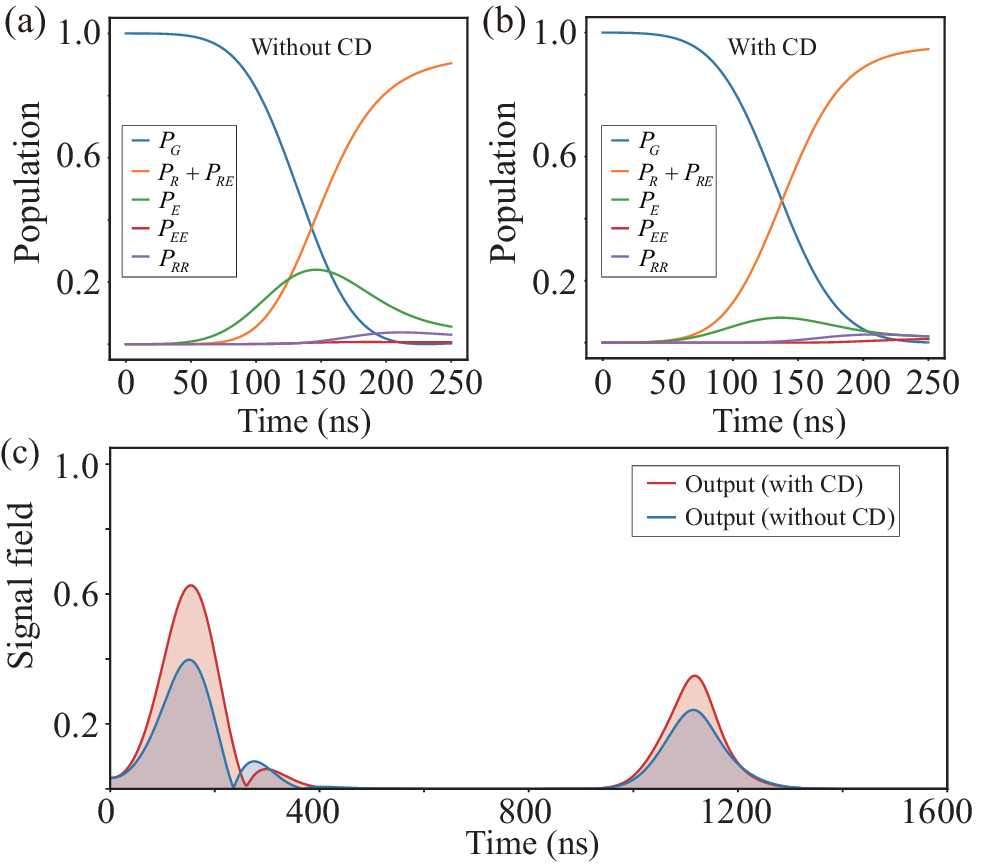}
	\caption{Population dynamics and output signal fields under multiphoton excitation ($\epsilon=0.3$) and imperfect Rydberg blockade ($U_{rr}/2\pi = 5$ MHz). (a) and (b) compare the atomic populations without and with CD driving, respectively. (c) illustrates the spatial propagation and retrieval of the signal fields. Other parameters are the same with Fig.~\ref{fig3}.}\label{fig7}
\end{figure}

To thoroughly evaluate the robustness of our scheme, we investigate the storage performance under non-ideal single-photon input and imperfect Rydberg blockade conditions. By introducing a weak-field amplitude correction factor $\epsilon = 0.3$, we account for the probability of secondary excitations induced by the multiphoton components in a weak coherent pulse. Additionally, a finite van der Waals interaction $U_{rr}/2\pi = 5$ MHz is set to simulate the imperfect Rydberg blockade effect. As shown in Fig.~\hyperref[fig7]{7(a)}, when the pulse duration is significantly shortened, the adiabatic condition is no longer satisfied in the absence of CD driving. Consequently, the increase in the lossy intermediate state $P_E$ reduces the final Rydberg excitation probability $(P_R + P_{RE})$ to $90.4\%$. In addition, a small but observable leakage appears in the double Rydberg state $P_{RR}$. In contrast, Fig.~\hyperref[fig7]{7(b)} shows that the introduction of CD driving effectively compensates for these non-adiabatic effects. The population in the intermediate state is strongly suppressed, and the final target Rydberg excitation is improved to $94.6\%$.

To explicitly simulate the macroscopic retrieval process, we numerically solve the expanded Lindblad master equation for the $6 \times 6$ density matrix $\rho(z,t)$
\begin{equation}
\frac{\partial \rho}{\partial t} = -i[H_T, \rho] + \mathcal{L}(\rho),
\end{equation}
where $H_T$ is the effective Hamiltonian defined above, and $\mathcal{L}(\rho)$ describes the dissipative dynamics of the system. The macroscopic propagation of the signal field is then evaluated by simultaneously solving the standard propagation equation of Eq.~(\ref{eq:MSE}), which is driven by the principal atomic coherence $\rho_{eg}(z,t) = \langle E|\rho(z,t)|G\rangle$ extracted directly from the expanded density matrix. Figure~\hyperref[fig7]{7(c)} shows the propagation of the signal field for the cases with and without CD driving in the presence of multiphoton input and imperfect Rydberg blockade. Without CD driving, the rapid evolution of the short writing pulse violates the adiabatic condition of the system. As a result, the population is transferred to the lossy intermediate state $P_E$, leading to strong spontaneous-emission scattering. Consequently, the input signal experiences significant absorption during the writing process, manifested as reduced transmitted leakage, and ultimately a relatively weak retrieved signal. By contrast, the proposed CD driving scheme compensates for these non-adiabatic effects and strongly suppresses the population of the intermediate state. This preserves the coherence of the storage process and enables much stronger and more efficient signal retrieval compared to the conventional EIT protocol without CD driving. Overall, our results demonstrate that the CD-assisted protocol possesses strong scalability and broad applicability, indicating that it can be extended to larger atomic ensembles and a variety of experimental settings.

\section{summary and outlook}\label{sec4}

In summary, we have proposed an STA protocol to accelerate the writing process of signal photons in a Rydberg-superatom-based EIT storage scheme. By engineering an appropriate CD driving field, the non-adiabatic excitation of the lossy intermediate state induced by rapid temporal modulation can be strongly suppressed. As a result, photonic excitations can be efficiently mapped onto the collective Rydberg state on timescales much shorter than those required by conventional adiabatic EIT protocols. By analyzing both the dynamics of the atomic population and the propagation of the signal field, we show that the proposed CD-assisted scheme significantly improves the writing performance and enables efficient retrieval of the stored excitation. The results further demonstrate that the protocol maintains robust performance against variations in the input signal field. Moreover, we analyze the influence of realistic control imperfections in CD driving. Even with Rabi-amplitude noise, phase errors, and off-resonant scattering, the system dynamics closely follow the ideal evolution, and the retrieved signal is almost unaffected. Furthermore, the scheme remains effective under experimentally relevant nonideal conditions such as multiphoton input and imperfect Rydberg blockade. Even when residual coupling to double Rydberg excitations is present, the CD-assisted protocol continues to suppress the unwanted population of the lossy intermediate state and maintains a clear advantage in storage and retrieval performance compared to the conventional EIT protocol without CD driving. These findings indicate that CD driving provides an efficient way to overcome the intrinsic speed limitation of adiabatic EIT storage while preserving high-fidelity mapping between the photonic excitation and the collective Rydberg spin wave. Therefore, our work offers a promising strategy for realizing fast and robust quantum memories based on Rydberg ensembles and may also be applicable to other quantum interfaces where rapid and coherent photonic state transfer is required.

\section{Acknowledgement}
This work was supported by the National Natural Science Foundation (Grant No. 12174048). W.L. acknowledges support from the EPSRC through Grant No. EP/W015641/1.

\section{DATA AVAILABILITY}
The data that support the findings of this study are openly
available \cite{wei2026}, embargo periods may apply.

\appendix
\section{Derivation of the counterdiabatic Hamiltonian}\label{App:A}

Consider a time-dependent Hamiltonian $H_0(t)$ with instantaneous non-degenerate eigenstates
\begin{equation}
H_0(t)\,|n(t)\rangle = E_n(t)\,|n(t)\rangle,\qquad \langle n(t)|m(t)\rangle=\delta_{nm},
\end{equation}
and we set $\hbar=1$. The goal of transitionless driving is to find an auxiliary Hamiltonian $H_{\mathrm{CD}}(t)$ such that the total Hamiltonian $H(t)=H_0(t)+H_{\mathrm{CD}}(t)$ admits solutions of the form
\begin{equation}
|\psi_n(t)\rangle = e^{i\alpha_n(t)}|n(t)\rangle,
\end{equation}
where $\alpha_n(t)$ incorporates both the dynamical and geometric phases accumulated during the evolution. That is, the system follows the instantaneous eigenstate $|n(t)\rangle$ exactly (up to this phase factor) without transitions to other eigenstates.

Substituting $|\psi_n(t)\rangle$ into the Schr\"odinger equation $i\partial_t|\psi_n\rangle = H|\psi_n\rangle$ and projecting onto the basis $\{|m(t)\rangle\}$ yields the condition for $m\neq n$:
\begin{equation}
\langle m|H_{\mathrm{CD}}|n\rangle = i\langle m|\partial_t n\rangle.
\end{equation}
One convenient closed form that satisfies these conditions is
\begin{equation}
\label{eq:Hcd_general}
H_{\mathrm{CD}}(t) = i\sum_n \Big(|\partial_t n(t)\rangle\langle n(t)| - \langle n(t)|\partial_t n(t)\rangle |n(t)\rangle\langle n(t)|\Big),
\end{equation}
where the second term accounts for the diagonal (Berry) phases and can be treated as a gauge choice. The off-diagonal components in Eq.~\eqref{eq:Hcd_general} cancel the non-adiabatic couplings $\langle m|\partial_t n\rangle$ that would otherwise induce transitions.

We now specialize in the resonant three-level ladder system considered
in the main text. In the symmetric single-excitation manifold, the
effective Hamiltonian reads
\begin{equation}
H_0(t) = -\frac{1}{2} \left[ \sqrt{N}\Omega_p(t) |E\rangle\langle G| + \Omega_c(t) |R\rangle\langle E|\right] + \text{H.c.} 
\end{equation}
with time-dependent probe and control Rabi frequencies $\sqrt{N}\Omega_p(t)$ and
$\Omega_c(t)$. 
The instantaneous dark state is given by
\begin{equation}
    |\Psi_0(t)\rangle = \cos\theta(t)|G\rangle - \sin\theta(t)|R\rangle.
\end{equation}
Straightforward differentiation yields $\partial_t|\Psi_0(t)\rangle = -\dot{\theta}(t)|B(t)\rangle$, where $|B(t)\rangle = \sin\theta(t)|G\rangle + \cos\theta(t)|R\rangle$ is the orthogonal state. Inserting this into Eq.~\eqref{eq:Hcd_general} to cancel the non-adiabatic couplings, we obtain the transitionless driving Hamiltonian:
\begin{equation}
    H_{\mathrm{CD}}(t) = i\big(|\partial_t \Psi_0\rangle\langle \Psi_0| - \mathrm{H.c.}\big) = i\dot{\theta}(t)\big(|\Psi_0\rangle\langle B| - |B\rangle\langle \Psi_0|\big).
\end{equation}
Expanding these states in the $\{|G\rangle, |R\rangle\}$ basis, the diagonal terms exactly cancel out, leading to the compact form
\begin{equation}
H_{\mathrm{CD}}(t) = i\frac{\Omega_{\mathrm{CD}}(t)}{2}\big(|G\rangle\langle R| - |R\rangle\langle G|\big)
\end{equation}
with
\begin{align}
    \Omega_{\mathrm{CD}}(t) = 
    \frac{\Omega_c(t)[\sqrt{N}\dot{\Omega}_p(t)] - [\sqrt{N}\Omega_p(t)]\dot{\Omega}_c(t)}
    {N\Omega_p^2(t) + \Omega_c^2(t)}.
\end{align}

\section{Derivation of the optical Bloch equations with wave-vector dependence}
\label{App:B}

We consider a ladder-type three-level atomic system $|g\rangle \leftrightarrow |e\rangle \leftrightarrow |r\rangle$ under the Rydberg blockade condition, such that the ensemble can be described by a collective superatom model. For simplicity, we assume that all electromagnetic fields propagate along the $z$-axis, allowing us to adopt a one-dimensional description of the spatial phase.

The probe field couples the $|G\rangle \leftrightarrow |E\rangle$ transition with a collectively enhanced Rabi frequency $\sqrt{N}\Omega_p$, whereas the control field couples the $|E\rangle \leftrightarrow |R\rangle$ transition with a Rabi frequency $\Omega_c$. An additional CD field directly couples $|G\rangle \leftrightarrow |R\rangle$ to suppress non-adiabatic transitions. The electric fields are expressed in the form of plane waves as:
\begin{align}
\Omega_p(z,t) &= \Omega_p(t) e^{i(k_p z - \omega_p t)}, \nonumber \\
\Omega_c(z,t) &= \Omega_c(t) e^{i(-k_c z - \omega_c t)}, \nonumber \\
\Omega_{\mathrm{CD}}(z,t) &= \Omega_{\mathrm{CD}}(t) e^{i(k_{\mathrm{CD}}z - \omega_{\mathrm{CD}} t)},
\end{align}
where $k_{p,c}$ and $\omega_{p,c}$ are the wave vectors and frequencies of the respective fields. Both the original two-photon excitation and the CD driving address the same collective spin-wave mode, characterized by the wave vector $k_{\mathrm{CD}}=k_p'-k_c'$.

Under the rotating-wave approximation and two-photon resonance conditions, the interaction Hamiltonian reads
\begin{align}
H_I =& -\frac{1}{2}\Big[
\sqrt{N}\Omega_p e^{i k_p z}\ket{E}\bra{G} + \Omega_c e^{-i k_c z}\ket{R}\bra{E} \\ \notag
&+ i \Omega_{\mathrm{CD}} e^{ik_{\mathrm{CD}}z}\ket{R}\bra{G}\Big] + \mathrm{H.c.}
\end{align}

The density matrix obeys the master equation
\begin{align}
\dot{\rho} = -{i}[H_I,\rho] + \mathcal{L}[\rho],
\end{align}
where $\mathcal{L}[\rho]$ accounts for the spontaneous decay of the intermediate state with a rate of $\Gamma$, and the decay of the Rydberg state occurs with a rate of $\Gamma_r$. Evaluating the commutator explicitly, one obtains the equations of motion with explicit spatial dependence:
\begin{align}
\frac{\partial}{\partial t}\tilde{\rho}_{rg} =& \frac{i}{2}\Omega_c e^{-i k_c z}\tilde{\rho}_{eg}
-\frac{\Gamma_r}{2}\tilde{\rho}_{rg}- \frac{i}{2} \sqrt{N} \Omega_p  e^{i k_p z} \tilde{\rho}_{re} \nonumber \\
&-\frac{1}{2}\Omega_{\mathrm{CD}} e^{ik_{\mathrm{CD}}z}(\rho_{gg}-\rho_{rr}), \nonumber \\
\frac{\partial}{\partial t}\tilde{\rho}_{eg} =&\frac{i}{2}\sqrt{N}\Omega_p e^{i k_p z}(\rho_{gg}-\rho_{ee})
+\frac{i}{2}\Omega_c e^{i k_c z}\tilde{\rho}_{rg} \nonumber \\ 
&-\frac{\Gamma}{2}\tilde{\rho}_{eg}+\frac{1}{2}\Omega_{\mathrm{CD}} e^{ik_{\mathrm{CD}}z}\tilde{\rho}_{er}, \nonumber \\
\frac{\partial}{\partial t}\tilde{\rho}_{er} =&\frac{i}{2}\sqrt{N}\Omega_p e^{i k_p z}\tilde{\rho}_{rg}^*
+\frac{i}{2}\Omega_c e^{i k_c z}(\rho_{rr}-\rho_{ee}) \nonumber \\ 
&- \frac{\Gamma+\Gamma_r}{2}\tilde{\rho}_{er}-\frac{1}{2}\Omega_{\mathrm{CD}} e^{-ik_{\mathrm{CD}}z}\tilde{\rho}_{eg}, \nonumber \\
\frac{\partial}{\partial t}\rho_{rr} =&\frac{i}{2}\Omega_c e^{-i k_c z}\tilde{\rho}_{er}
-\frac{i}{2}\Omega_c^* e^{i k_c z}\tilde{\rho}_{er}^*-\Gamma_r\rho_{rr}  \nonumber \\
&-\frac{1}{2} \Omega_{\mathrm{CD}} (\tilde{\rho}_{rg} e^{-i k_{\mathrm{CD}} z}
+ \tilde{\rho}_{rg}^* e^{i k_{\mathrm{CD}} z}), \nonumber \\
\frac{\partial}{\partial t}\rho_{ee} =&-\frac{i}{2}\sqrt{N}\Omega_p^* e^{-i k_p z}\tilde{\rho}_{eg}
+\frac{i}{2}\sqrt{N}\Omega_p e^{i k_p z}\tilde{\rho}_{eg}^* \nonumber \\
&-\frac{i}{2}\Omega_c e^{-i k_c z}\tilde{\rho}_{er}
+\frac{i}{2}\Omega_c^* e^{i k_c z}\tilde{\rho}_{er}^*
-\Gamma\rho_{ee}, \nonumber \\
\frac{\partial}{\partial t}\rho_{gg} =&\frac{i}{2}\sqrt{N}\Omega_p^* e^{-i k_p z}\tilde{\rho}_{eg}
-\frac{i}{2}\sqrt{N}\Omega_p e^{i k_p z}\tilde{\rho}_{eg}^*+\Gamma\rho_{ee} \nonumber \\
&+\Gamma_r\rho_{rr}+\frac{1}{2} \Omega_{\mathrm{CD}} (\tilde{\rho}_{rg} e^{-i k_{\mathrm{CD}} z}
+ \tilde{\rho}_{rg}^* e^{i k_{\mathrm{CD}} z}).
\end{align}

To remove the explicit spatial component, we introduce the slowly varying coherences:
\begin{align}
\rho_{eg}(z,t) &= \tilde{\rho}_{eg}(t)\, e^{i k_p z}, \nonumber \\
\rho_{er}(z,t) &= \tilde{\rho}_{er}(t)\, e^{i k_c z}, \nonumber \\
\rho_{rg}(z,t) &= \tilde{\rho}_{rg}(t)\, e^{i (k_p-k_c) z}.
\end{align}
Substituting these expressions into the equations above, all spatial
phase factors cancel identically.
As a result, the equations for the slowly varying envelopes are:
\begin{align}
\frac{\partial}{\partial t}\rho_{rg} =& \frac{i}{2}\Omega_c \rho_{eg} -  \frac{\Gamma_r}{2}\rho_{rg} -  \frac{i}{2} \sqrt{N} \Omega_p \rho_{re}\nonumber \\
&- \frac{1}{2}\Omega_{\mathrm{CD}} e^{i \Delta k z}(\rho_{gg} - \rho_{rr}), \nonumber \\
\frac{\partial}{\partial t}\rho_{eg} =& \frac{i}{2}\sqrt{N}\Omega_p (\rho_{gg} - \rho_{ee}) + \frac{i}{2}\Omega_c \rho_{rg} - \frac{\Gamma}{2}\rho_{eg} \nonumber \\
&+ \frac{1}{2}\Omega_{\mathrm{CD}} e^{i \Delta k z}\rho_{er}, \nonumber \\
\frac{\partial}{\partial t}\rho_{er} =& \frac{i}{2}\sqrt{N}\Omega_p \rho_{rg}^* + \frac{i}{2}\Omega_c (\rho_{rr} - \rho_{ee}) - \frac{\Gamma+\Gamma_r}{2}\rho_{er} \nonumber \\
&- \frac{1}{2}\Omega_{\mathrm{CD}} e^{-i \Delta k z}\rho_{eg}, \nonumber \\
\frac{\partial}{\partial t}\rho_{rr} =& \frac{i}{2}\Omega_c \rho_{er} - \frac{i}{2}\Omega_c^* \rho_{er}^* - \Gamma_r \rho_{rr} \nonumber \\
&-\frac{1}{2} \Omega_{\mathrm{CD}} (\rho_{rg} e^{-i \Delta k z}
+ \rho_{rg}^* e^{i \Delta k z}), \nonumber \\
\frac{\partial}{\partial t}\rho_{ee} =& -\frac{i}{2}\sqrt{N}\Omega_p^* \rho_{eg} + \frac{i}{2}\sqrt{N}\Omega_p \rho_{eg}^* \nonumber  \nonumber \\
&- \frac{i}{2}\Omega_c \rho_{er} + \frac{i}{2}\Omega_c^* \rho_{er}^* - \Gamma \rho_{ee}, \nonumber \\
\frac{\partial}{\partial t}\rho_{gg} =& \frac{i}{2}\sqrt{N}\Omega_p^* \rho_{eg} - \frac{i}{2}\sqrt{N}\Omega_p \rho_{eg}^* + \Gamma \rho_{ee}+\Gamma_r\rho_{rr}\nonumber \\
&+\frac{1}{2} \Omega_{\mathrm{CD}} (\rho_{rg} e^{-i \Delta k z}
+ \rho_{rg}^* e^{i \Delta k z}),
\end{align}
where $\Delta k=|k_p-k_c-k_{\mathrm{CD}}|$.

\section{Derivation of the Maxwell--Schrödinger Equation in the Collective Basis}\label{App:C}

In this Appendix, we present a detailed derivation of the MSE employed in the main text, with particular emphasis on the origin of the collective enhancement factor $\sqrt{N}$ in the atom--field coupling. The derivation follows standard quantum-optical treatments and highlights the distinction between the physical polarization of the medium and the collective variables introduced to describe symmetric single-excitation dynamics in the ensemble.

Under the slowly varying envelope approximation, the propagation of the probe field envelope $\Omega_p(z,t)$ along the $z$ direction is governed by~\cite{PhysRevA.69.043801}
\begin{equation}
\left( \frac{\partial}{\partial z} + \frac{1}{c}\frac{\partial}{\partial t} \right)\Omega_p(z,t)
= i \frac{\omega_p}{2\varepsilon_0 c\hbar} d_{eg} \, P(z,t),
\label{eq:Maxwell_general}
\end{equation}
where $\omega_p$ is the probe frequency, $c$ is the speed of light in vacuum, $d_{eg}$ is the electric dipole matrix element of the $|g\rangle \leftrightarrow |e\rangle$ transition, and $P(z,t)$ denotes the macroscopic polarization density of the atomic medium.

The physical polarization is defined as the sum of the dipole moments of all atoms within a coarse-grained volume,
\begin{equation}
P(z,t)
= \frac{1}{V}\sum_{j=1}^{N} d_{eg}\,
\langle \hat{\sigma}^{(j)}_{ge}(z,t) \rangle ,
\label{eq:physical_polarization}
\end{equation}
where $\hat{\sigma}^{(j)}_{ge}=|g_j\rangle\langle e_j|$ is the lowering operator of the $j$th atom and $V$ is the quantization volume (typically the Rydberg blockade volume) containing the $N$ atoms. We restrict the dynamics to the single-excitation manifold, where at most one atom is excited from $\ket{g}$ to $\ket{e}$ within the ensemble. In this subspace, it is convenient to introduce the normalized collective lowering operator
\begin{equation}
\hat{S}_{ge}
= \frac{1}{\sqrt{N}} \sum_{j=1}^{N} \hat{\sigma}^{(j)}_{ge},
\label{eq:collective_operator}
\end{equation}
whose expectation value defines the collective coherence
\begin{equation}
\rho_{eg}
\equiv \langle \hat{S}_{ge} \rangle .
\end{equation}

Using Eqs.~(\ref{eq:physical_polarization}) and (\ref{eq:collective_operator}), the physical polarization can be expressed as
\begin{equation}
P(z,t)
= \frac{\sqrt{N}}{V} d_{eg}\, \rho_{eg}(z,t).
\label{eq:polarization_superatom}
\end{equation}
This relation shows explicitly that the collective variable $\rho_{eg}$ differs from the physical polarization by a factor of $\sqrt{N}$.

Substituting Eq.~(\ref{eq:polarization_superatom}) into 
Eq.~(\ref{eq:Maxwell_general}) gives
\begin{equation}
\left( \partial_z + \frac{1}{c}\partial_t \right)\Omega_p=i \frac{\omega_p d_{eg}^2}{2\varepsilon_0 c\hbar}
\frac{\sqrt{N}}{V}\,\rho_{eg}.
\end{equation}
Using the atomic number density $n = N/V$, we rewrite
\begin{equation}
\frac{\sqrt{N}}{V} = \frac{n}{\sqrt{N}},
\end{equation}
so that
\begin{equation}
\left( \partial_z + \frac{1}{c}\partial_t \right)\Omega_p=i \frac{\omega_p d_{eg}^2}{2\varepsilon_0 c\hbar}
\frac{n}{\sqrt{N}}\,\rho_{eg}.
\end{equation}
Introducing the resonant optical depth
\begin{equation}
\alpha=\frac{n d_{eg}^2 \omega_p L}{\varepsilon_0 \hbar c \Gamma},
\end{equation}
where $L$ is the length of the medium and $\Gamma$ the spontaneous decay rate of the intermediate state $\ket{e}$, we use
\begin{equation}
\frac{\omega_p d_{eg}^2}{\varepsilon_0 \hbar}=\frac{\alpha \Gamma }{n L}
\end{equation}
to finally obtain
\begin{equation}
\left( \partial_z + \frac{1}{c}\partial_t \right)\Omega_p=i \frac{\alpha \Gamma }{2\sqrt{N} L}\,\rho_{eg}.
\label{eq:MSE_superatom}
\end{equation}

Equation~(\ref{eq:MSE_superatom}) is the Maxwell--Schrödinger equation used in the main text. The factor $1/\sqrt{N}$ in the denominator compensates for the collective enhancement already incorporated in the definition of superatom coherence $\rho_{eg}$. Consequently, the propagation equation remains consistent with the physical polarization of the medium and avoids double counting of the collective enhancement.

\section{Physical Implementation of the CD Field and AC Stark Shift Compensation}\label{App:D}

In the theoretical framework of CD driving for a ladder-type three level system, the CD field is phenomenologically introduced as a direct coupling between the ground state $|G\rangle$ and the Rydberg state $|R\rangle$. However, since the $|G\rangle \leftrightarrow |R\rangle$ dipole transition is practically forbidden, this effective coupling must be realized through a two-photon Raman process utilizing the intermediate excited state $|E\rangle$. 

To induce the effective CD coupling $\Omega_{\mathrm{CD}}(t)$, we introduce two auxiliary, large detuned laser fields $\Omega_p'(t)$ and $\Omega_c'(t)$ with wavevectors $k_p'$ and $k_c'$, respectively. Both auxiliary fields are red-detuned from the single-photon transition by a large amount $\Delta$ ($\Delta \gg \Omega_{p,c}', \Gamma$), acquiring a rapidly oscillating phase $e^{i\Delta t}$ in the rotating frame to maintain the two-photon resonance condition. The total light fields addressing the $|G\rangle \leftrightarrow |E\rangle$ and $|R\rangle \leftrightarrow |E\rangle$ transitions are the superposition of the resonant pulses and the off-resonant auxiliary Raman fields:
\begin{align}
    \tilde{\Omega}_p(z, t) &= \sqrt{N}\Omega_p(t) e^{i k_p z} + \Omega_p'(t) e^{i k_p' z} e^{i\Delta t}, \\
    \tilde{\Omega}_c(z, t) &= \Omega_c(t) e^{-i k_c z} + \Omega_c'(t) e^{-i k_c' z} e^{-i\Delta t}.
\end{align}

Under the rotating wave approximation, the time-dependent physical Hamiltonian reads:
\begin{equation}
    H_{\mathrm{n}} = -\frac{1}{2} \Big[ \tilde{\Omega}_p(z,t) |E\rangle\langle G| + \tilde{\Omega}_c(z,t) |R\rangle\langle E| \Big]+ \mathrm{H.c.} 
\end{equation}
Under the large-detuning condition, the rapidly oscillating terms $e^{\pm i\Delta t}$ do not produce a real single-photon population in the intermediate state. By adiabatically eliminating the intermediate state $|E\rangle$, the auxiliary fields give rise to an effective Raman coupling directly connecting $|G\rangle$ and $|R\rangle$, leading to the following effective Hamiltonian:
\begin{align}
    H_{\mathrm{eff}} =& -\frac{1}{2} \Big[ \sqrt{N}\Omega_p(t) e^{i k_p z} |E\rangle\langle G| + \Omega_c(t) e^{-i k_c z} |R\rangle\langle E| \nonumber \\
    &+ i \Omega_{\mathrm{CD}}(t) e^{i  k_\mathrm{CD} z} |R\rangle\langle G|  \Big]+ \mathrm{H.c.}
\end{align}
By matching the physical Raman coupling $\frac{\Omega_p'(t) {\Omega_c'^*(t)}}{4\Delta} e^{i(k_p' - k_c')z}$ to the target CD term $-i \frac{\Omega_{\mathrm{CD}}(t)}{2} e^{i k_\mathrm{CD} z}$, we set $k_p' - k_c' = k_\mathrm{CD}$ and establish the following amplitude mapping relations:
\begin{align}
    \Omega_p'(t) &= \sqrt{2\Delta |\Omega_{\mathrm{CD}}(t)|}, \\
    \Omega_c'(t) &= i \cdot \mathrm{sgn}(\Omega_{\mathrm{CD}}(t)) \sqrt{2\Delta |\Omega_{\mathrm{CD}}(t)|}.
\end{align}
While adiabatic elimination provides the desired cross-coupling, the intensely large detuned Raman fields inherently induce dynamical AC Stark shifts on the energy levels. For large detuned fields, the physical energy levels are shifted by:
\begin{align}
    \delta_G(t) &= \frac{|\Omega_p'(t)|^2}{4\Delta} = \frac{1}{2} |\Omega_{\mathrm{CD}}(t)|, \label{eq:stark_g} \\
    \delta_R(t) &= \frac{|\Omega_c'(t)|^2}{4\Delta} = \frac{1}{2} |\Omega_{\mathrm{CD}}(t)|, \label{eq:stark_r} \\
    \delta_E(t) &= -\frac{|\Omega_p'(t)|^2 + |\Omega_c'(t)|^2}{4\Delta} = - |\Omega_{\mathrm{CD}}(t)|. \label{eq:stark_e}
\end{align}
To prevent these dynamical Stark shifts from disrupting the resonance conditions, they can be completely canceled out. In practice, this can be achieved experimentally by introducing additional auxiliary compensation laser fields or by applying real-time frequency modulations. Accordingly, we introduce a compensation Hamiltonian $H_{\mathrm{s}}$ to exactly negate these physical energy shifts:
\begin{equation}
    H_{\mathrm{s}} = -\frac{1}{2} |\Omega_{\mathrm{CD}}(t)| (|G\rangle\langle G| +|R\rangle\langle R|)+ |\Omega_{\mathrm{CD}}(t)| |E\rangle\langle E|.
\end{equation}
By incorporating the Raman fields alongside this strict AC Stark shift compensation, our model accurately captures the complete physical dynamics.

\begin{figure}
	\centering
	\includegraphics[width=1\linewidth]{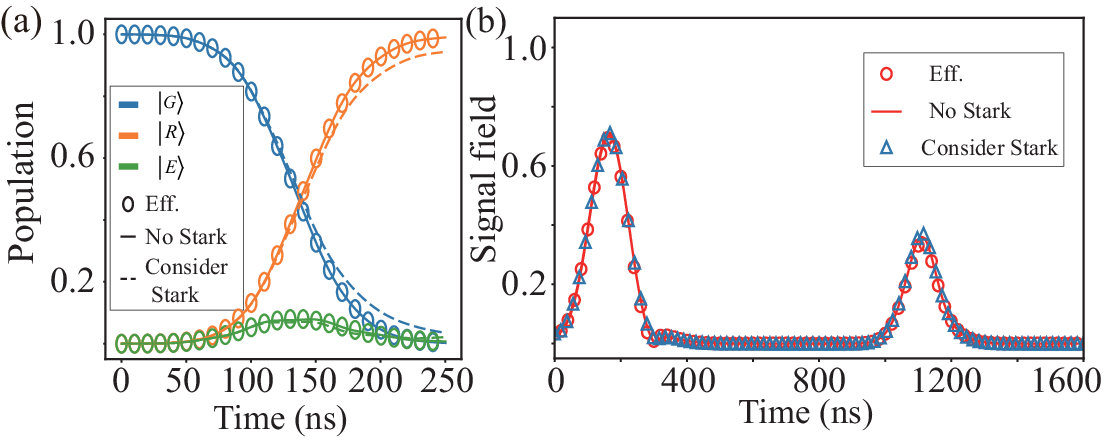}
	\caption{Population dynamics in (a) and output signal fields in (b) for the ideal effective CD model, the physical Raman model with AC Stark-shift, and the physical model without AC Stark-shift. The large detuning is $\Delta/2\pi = 10$ GHz, other parameters are the same with Fig.~\ref{fig3}.}\label{fig8}
\end{figure}

To numerically validate our theoretical framework, we simulate the population dynamics and the propagation of the signal field. Comparisons among the ideal effective CD model, the physical Raman model with AC Stark shift, and the physical model without AC Stark shift are presented in Figure~\hyperref[fig8]{8}. Figure~\hyperref[fig8]{8(a)} shows the population dynamics during the evolution. The dynamics of the physical model without Stark shift closely match those of the ideal effective CD model. In contrast, the physical model with Stark shift exhibits a slightly lower transfer fidelity. Despite this minor deviation in the populations, Figure~\hyperref[fig8]{8(b)} demonstrates that the propagation of the signal field remains highly robust. The retrieved signal pulses from all three models are nearly identical, indicating that the absence of AC Stark shift compensation has a negligible impact on the overall optical storage and retrieval. The underlying physical reason is that the intense Raman fields symmetrically shift the ground state $|G\rangle$ and the Rydberg state $|R\rangle$, inherently preserving the crucial two-photon resonance condition. Therefore, high-efficiency optical memory can still be achieved experimentally even without real-time frequency modulation to compensate for AC Stark shifts.

\bibliography{manuscript.bbl}

\begin{thebibliography}{90}
\bibitem [{pul()}]{pulse_parameters_2026}%
  \BibitemOpen
  \href@noop {} {}\bibinfo {note} {The control pulse is defined as $\Omega_c(t) = \Omega_{c}^{\mathrm{max}} \cdot \frac{1}{2}[1 + \mathrm{erf}(\frac{t - t_{\mathrm{on}}}{w_s})] \cdot \frac{1}{2}[1 + \mathrm{erf}(\frac{t_{\mathrm{off}} - t}{w_s})]$, where $t_{\mathrm{on}}=-0.8T$, $t_{\mathrm{off}}=0.69T$, and $w_s=T/6$.}\BibitemShut {Stop}%
\end{thebibliography}

\end{document}